\PassOptionsToPackage{pdfpagelabels=false}{hyperref} 
\documentclass{aa}

\usepackage{natbib}
\usepackage{graphicx}
\usepackage{txfonts}
\usepackage{amsmath}
\usepackage{amssymb}	
\usepackage{rotating}
\usepackage{natbib}

\bibpunct{(}{)}{;}{a}{}{,} 
\usepackage[colorlinks,linkcolor=blue,anchorcolor=blue,citecolor=blue]{hyperref}

\graphicspath{{./}{Figures/}}

\begin{document}


\title{The eROSITA Final Equatorial-Depth Survey (eFEDS)}
\subtitle{LOFAR view of brightest cluster galaxies and AGN feedback}
\titlerunning{LOFAR view of brightest cluster galaxies in eFEDS}

\author{T. Pasini\inst{\ref{inst1}} 
\and M. Brüggen\inst{\ref{inst1}} 
\and D. N. Hoang\inst{\ref{inst1}}
\and V. Ghirardini\inst{\ref{inst2}}
\and E. Bulbul\inst{\ref{inst2}}
\and M. Klein\inst{\ref{inst3}}
\and A. Liu\inst{\ref{inst2}}
\and T. W. Shimwell\inst{\ref{inst4},\ref{inst5}}
\and M. J. Hardcastle\inst{\ref{inst6}}
\and W. L. Williams\inst{\ref{inst5}}
\and A. Botteon\inst{\ref{inst5}}
\and F. Gastaldello\inst{\ref{inst7}}
\and R. J. van Weeren\inst{\ref{inst5}}
\and A. Merloni\inst{\ref{inst2}}
\and F. de Gasperin\inst{\ref{inst1}, \ref{inst8}} 
\and Y. E. Bahar\inst{\ref{inst2}}
\and F. Pacaud\inst{\ref{inst9}}
\and M. Ramos-Ceja\inst{\ref{inst2}}
}

\authorrunning{T. Pasini, M. Brüggen, D. N. Hoang et al.}

\institute{Hamburger Sternwarte, Universität Hamburg, Gojenbergsweg 112, 21029 Hamburg, Germany\label{inst1}
\and Max Planck Institute for Extraterrestrial Physics, Giessenbachstrasse 1, 85748 Garching, Germany\label{inst2}
\and Faculty of Physics, Ludwig-Maximilians-Universit{\"a}t, Scheinerstr. 1, 81679, Munich, Germany\label{inst3}
\and ASTRON, The Netherlands Institute for Radio Astronomy, Postbus 2, 7990 AA Dwingeloo, The Netherlands\label{inst4}
\and Leiden Observatory, Leiden University, P.O. Box 9513, 2300 RA Leiden, The Netherlands\label{inst5}
\and Centre for Astrophysics Research, School of Physics, Astronomy and Mathematics, University of Hertfordshire, College Lane, Hat- field AL10 9AB, UK\label{inst6}
\and IASF – Milano, INAF, Via A. Corti 12, I-20133 Milano, Italy\label{inst7}
\and INAF - Istituto di Radioastronomia, via P. Gobetti 101, 40129, Bologna, Italy\label{inst8}
\and Argelander-Institut f\"ur Astronomie (AIfA), Universit\"at Bonn, Auf dem H\"ugel 71, 53121 Bonn, Germany\label{inst9}
}


\abstract
   {During the performance verification phase of the SRG/eROSITA telescope, the eROSITA Final Equatorial-Depth Survey (eFEDS) has been carried out. It covers a 140 deg$^2$ field located at 126$^\circ <$ R.A. $< 146^\circ$ and -3$^\circ <$ Dec. $< +6^\circ$ with a nominal unvignetted exposure over the field of 2.2 ks. 542 candidate clusters and groups were detected in this field, down to a flux limit $F_X \sim 10^{-14}$ erg s$^{-1}$ cm$^{-2}$ in the 0.5-2 keV band. 
   }
   {In order to understand radio-mode feedback in galaxy clusters, we study the radio emission of brightest cluster galaxies of eFEDS clusters and groups, and we relate it to the X-ray properties of the host cluster. }
   {Using LOFAR we identify 227 radio galaxies hosted in the BCGs of the 542 galaxy clusters and groups detected in eFEDS. We treat non-detections as radio upper limits. We analyse the properties of radio galaxies, such as redshift and luminosity distribution, offset from the cluster centre, largest linear size and radio power. We study their relation to the intracluster medium of the host cluster. 
   }
   {We find that BCGs with radio-loud AGN are more likely to lie close to the cluster centre than radio-quiet BCGs. There is a clear relation between the cluster's X-ray luminosity and the 144 MHz radio power of the BCG. Statistical tests indicate that this correlation is not produced by biases or selection effects in the radio band. We see no apparent link between largest linear size of the radio galaxy and the central density of the host cluster. Converting the radio luminosity to kinetic luminosity, we find that radiative losses of the intracluster medium are in an overall balance with the heating provided by the central AGN. Finally, we tentatively classify our objects into disturbed and relaxed based on different morphological parameters, and we show that the link between the AGN and the ICM apparently holds for both subsamples, regardless of the dynamical state of the cluster.}
   {}
   


\maketitle

\section{Introduction}
\label{sec:intro}

Radio galaxies that sit at the centres of galaxy clusters and galaxy groups play an important role in regulating the temperature of the Intra-Cluster Medium (ICM) and Intra-Group Medium (IGrM).  
Radio-loud Active Galactic Nuclei (AGN) are usually hosted by Brightest Cluster Galaxies (BCG) and they quench the cooling of the hot ($\sim$ 10$^7$ K) ICM through mechanical feedback (see e.g. reviews by \citealt{McNamara-Nulsen_2012, Gitti_2012}). 
Effects of AGN feedback are manifested in the form of X-ray cavities and ripples in the cluster atmosphere \citep[e.g.,][]{McNamara_2000, Birzan_2004, Fabian_2006, Markevitch_2007, Gastaldello_2009}. Consequences are also observed in the thermodynamical properties of the ICM, such as the gas entropy distribution \citep[e.g.,][]{Cavagnolo_2009}, or in the transport of high-metallicity gas from the cluster centre to the outskirts \citep[e.g.,][]{Liu_2019b}. This type of feedback is generally positive, in the sense that when the radiative losses of the ICM increase, the AGN counteracts this by heating the ICM. The more gas cools and fuels the Super-Massive Black Hole (SMBH) at the centre of the BCG, the higher the energy output that is able to quench the ICM radiative losses and establish what is commonly known as \textit{AGN feedback loop} (see review from \citealt{Gaspari_2020}). AGN feedback has been observed in systems ranging from isolated elliptical galaxies \citep{Croton_2006, Sijacki_2015, O'Sullivan_2011b} to massive clusters where it prevents the formation of cooling flows \citep{McDonald_2019, Ehlert_2011, Pasini_2021}.
Most of the AGN associated with BCGs are in the so-called ‘radio-mode' or 'maintenance-mode' (to distinguish it from the radiatively dominated quasar-mode feedback), where the accretion rate is modest and the feedback is mediated via mechanical work from powerful jets. A scaling relation between cavity power and radio luminosity, spanning over seven orders of magnitude in radio and jet power, has been observed in nearby systems \citep{Birzan_2004, Merloni_2007, Birzan_2008, Cavagnolo_2010, O'Sullivan_2011a, Heckman_2014}.

It has been pointed out that AGN feedback may operate differently in galaxy groups, where the gravitational potential is shallower \citep{Sun_2012}. Here, less energetic AGN than in clusters can have a larger impact on the IGrM \citep{Giodini_2010}, since outbursts are also capable of expelling cool gas from the central region \citep{Alexander_2010, Morganti_2013}.
As a result, AGN feedback may break the self-similarity between galaxy clusters and groups, especially in terms of their baryonic properties \citep{Jetha_2007}. 
Hence, galaxy groups may be particularly interesting to study AGN feedback because their different environment should be reflected in the properties of the central AGN \citep[e.g.,][]{Giacintucci_2011b}. 

\cite{vonderLinden_2007} found that Brightest Group Galaxies (BGGs) and BCGs lie on a different Fundamental Plane - in terms of velocity dispersion, effective radius and average surface brightness - and have experienced star formation for a shorter time than non-BCGs \footnote{hereafter we will refer to them as satellites for more clarity.}. In the companion paper by \citet{Best_2007}, they also argued that BCGs are more likely to host radio-loud AGN than satellites of the same mass (cluster-hosted and not), but are less likely to host an optical AGN. These differences are particularly pertinent for BGGs. \citet{Main_2017} studied the relation betweeen AGN feedback and central (at 0.004$R_{500}$) cooling time in a sample of 45 galaxy clusters. They find a clear correlation between AGN power and halo mass and X-ray luminosity in clusters with a central cooling time of $<$ 1 Gyr.

X-ray observations of galaxy groups are more difficult than for galaxy clusters since groups have lower surface brightnesses and emit at lower temperatures, outside of the sweet spot of most X-ray observatories (see e.g., \citealt{Willis_2005}).
Still there has been some notable work on groups. \citet{Lovisari_2015} have presented scaling relations in the group regime, while \citet{Johnson_2009} and \citet{O'Sullivan_2017} have classified their samples of groups into cool-core and non-cool-core. \citet{Kolokythas_2018} have focused on central radio galaxies in the so-called Complete Local Volume Group Sample (CLoGS), and found that $\sim$ 92\% of groups in their high-richness sample (26 objects) have dominant galaxies (BGGs) hosting radio sources. They also argued that radio galaxies showing jets are more common in bright groups, while radio non-detections are mostly found in X-ray faint systems. The same authors report, in the CLoGS low-richness sample (27 objects) studied in \citet{Kolokythas_2019}, a radio detection rate of $\sim$82\% in the luminosity range $10^{20} - 10^{25}$ W Hz$^{-1}$ at 235 MHz.
\citet{Malarecki_2015} proposed that the lower densities in the IGrM, compared to the ICM, allows the lobes of group radio galaxies to expand to large distances. \cite{Werner_2014} used Far InfraRed (FIR), optical, and X-ray data to study eight nearby giant elliptical galaxies, all central members of relatively low-mass groups. They find evidence that cold gas in those centrals galaxies is produced mostly by cooling from the hot phase and that this cool gas fuels outbursts of the AGN. \citet{Dunn_2010} investigated a statistically complete sample of 18 nearby massive galaxies with X-ray and radio coverage, finding that 10 of them exhibit extended radio emission, with 9 also showing hints of interplay with the surrounding hot gas.

\citet{mhr09} determined that all cool-core clusters in a complete sample of $\sim 60$ clusters show a central radio galaxy, while only half of non-cool core clusters have one. Interestingly, when extending this study to galaxy groups, the trend becomes much weaker \citep{brs14,brl15}. 
A similar result was recently discussed in \citet{Pasini_2020} (hereafter P20). In this paper, the authors studied a sample of 247 X-ray detected galaxy groups in the COSMOS field, matching them to radio galaxies detected in the VLA-COSMOS Deep Survey \citep{Schinnerer_2010} and in the COSMOS MeerKAT survey (MIGHTEE, \citealt{Jarvis_2016}). They found that more than 70\% of their radio galaxies are not hosted in BGGs, while in clusters $\sim$85\% of central radio galaxies are associated with BCGs. They also discuss a correlation between the X-ray luminosity of groups and the radio power from the central radio galaxy since more massive groups seem to host more powerful sources. Indeed, \citet{Pasini_2021b} recently showed that, in their sample of groups, BGGs showing powerful radio emission are always found within 0.2 $R_{\rm vir} \sim 0.3 R_{200}$ from the centre. 

The extended ROentgen Survey with an Imaging Telescope Array (eROSITA) onboard the Spectrum-Roentgen-Gamma (SRG) mission \citep{Predehl_2021} was launched on July 13$^{\rm }$, 2019. The large effective area (1365 cm$^2$ at 1 keV), large field of view (FoV, 1~deg diameter), good spatial resolution (half energy width of 26 $\arcsec$ averaged over the FoV at 1.49~keV, 16 $\arcsec$ on-axis) and spectral resolution ($\sim 80$~eV full width half maximum at 1~keV) of eROSITA allow unique survey science capabilities by scanning large areas of the X-ray sky quickly and efficiently \citep{Merloni_2012}. Thus, eROSITA is detecting a large number of previously undetected groups and clusters, most of them with low surface brightnesses and at low redshifts, even though the confirmation of these groups in the optical is challenging for $z<0.1-0.2$.

In this work, we exploit the results of the eROSITA Final Equatorial-Depth Survey (eFEDS), a mini-survey designed to demonstrate the science capabilities of eROSITA. We study the radio galaxies observed in cluster centres at a frequency of 144 MHz by the LOw Frequency ARray (LOFAR, \citealt{vanHaarlem_2013}) in order to investigate their relation to their host clusters. This paper is structured as follows: in Sec.~\ref{sec:sample} we give a detailed description of how we build the sample. In Sec.~\ref{sec:analysis} we show the results, compare them to previous work and analyse the implications for AGN feedback. Finally, in Sec.~\ref{sec:conclusions} we summarise our results. Throughout this paper, we assume a standard $\Lambda$CDM cosmology with H$_0 = 70$ km s$^{-1}$ Mpc$^{-1}$, $\Omega_\Lambda = 0.7$ and $\Omega_{\text{M}} =  1-\Omega_\Lambda  = 0.3$.


\section{The sample}
\label{sec:sample}

\subsection{The eROSITA observation of eFEDS and the cluster catalog}
\label{sec:efeds}

eFEDS covers a 140 deg$^2$ field located in an equatorial region, with R.A. from $\sim$126 to $\sim$146 deg, and declination from $\sim$-3 to $\sim$6 deg. This field was uniformly scanned by eROSITA during the Performance Verification phase resulting in a nominal exposure of about 2.2 ks (unvignetted) over the field, which is similar in depth to the final exposure that will be reached in 4 years in equatorial fields in the eROSITA All-sky survey \citep{LiuAng_2021}.

The eFEDS data were acquired by eROSITA over 4 days, between November 4 and 7, 2019. These data were processed by the eROSITA Standard Analysis Software System (eSASS, \citealt{Brunner_2021}). We refer to \citet[][hereafter G21]{Ghirardini_2021} for further details on the data processing.
The source detection was performed using the tool {\tt erbox} in eSASS, on the merged 0.2 – 2.3 keV image of all seven eROSITA Telescope Modules (TMs). {\tt erbox} is a modified sliding box algorithm, which searches for sources in the input image that are brighter than the expected background fluctuation at a given image position. For each candidate source, the detection likelihood and the extent likelihood are determined by fitting the image with the source model, which is a $\beta$-model convolved with the calibrated PSF. Sources with extension too broad to be fitted by the PSF have a larger extent likelihood. For further details on the source detection procedure we refer to \citet{Brunner_2021}. We detect 542 candidate clusters over the full field \citep{LiuAng_2021}. This corresponds to a source density of $\sim$ 4 clusters per square degree at the equatorial depth. Photometric redshifts are obtained through the Multi-Component Matched Filter (MCMF) cluster confirmation tool \citep{Klein_2018}. Optical data from the Hyper Suprime-Cam Subaru Strategic Program (HSC-SSP, \citealt{Aihara_2018}) and from the DESI Legacy Survey (LS, \citealt{Dey_2019}) were exploited. We refer to \citet{Klein_2021} and \citet{LiuAng_2021} for further details. Spectroscopic redshifts derived from 2MRS \citep{Huchra_2012}, SDSS \citep{Blanton_2017} or GAMA \citep{Driver_2009} are used when available (296 out of 542 clusters). For each cluster, a massive red-sequence galaxy near the X-ray emission peak is selected as the BCG, following \citet{Klein_2021}.

The sample of clusters can be expected to be contaminated by spurious sources, or misclassified AGN, at a level of $\sim$19.7\% (see \citealt{LiuAng_2021}). Cluster contamination is therefore taken into account through the parameter $f_{\mathrm{cont},i}$ \citep{Klein_2019}, which is defined as:

\begin{equation}
   f_{\mathrm{cont},i}=\frac{\int_{\lambda_i}^{\infty} f_\mathrm{rand}(\lambda,z_i) d\lambda}{\int_{\lambda_i}^{\infty}
f_\mathrm{obs}(\lambda,z_i) d\lambda},
\label{eq:fcont}
\end{equation}
where $f_\mathrm{rand,z}$ is the richness distribution of random positions at the cluster candidate redshift $z_i$, $f_\mathrm{obs}(\lambda,z_i)$ the richness distribution of true candidates and $\lambda_i$ the richness of the cluster candidate. The estimator $f_{\mathrm{cont},i}$ is correlated with the probability of a source being a chance superposition. Applying a given cut to this parameter allows us to reduce the initial contamination of the cluster sample. Assuming independence between contaminants in the X-ray sample, the fractional contamination of the sample is simply the product of the initial fractional contamination of the X-ray sample and the applied cut in $f_\mathrm{cont}$. For example, applying a cut in $f_{\mathrm{cont},i} < 0.3$ results in a sample of about 88\% (477/542) of the eFEDS extended sources being confirmed as galaxy clusters. Assuming an initial contamination of the X-ray sample of 20\% \citep{Klein_2021}, this $f_\mathrm{cont}$ selected sample is expected to contain $\sim$6\% contamination. Subsequent tests described in \citet{Klein_2021} confirm the expected amount of contamination to be $6\pm3$\%. For more details about the X-ray catalog we refer to \citet{LiuAng_2021}, while further details on the optical confirmation and contamination can be found in \citet{Klein_2021}.


\subsection{The LOFAR observations of eFEDS and the radio source catalog}

\begin{table}
	\centering
	\caption{LOFAR HBA observations of the eFEDS field}
	\begin{tabular}{lc}
		\hline\hline
		Telescope                  &    LOFAR                  \\ \hline
		Project                       &    LC13\_029, LT5\_007, \\
		           &             LT10\_010, LT14\_004             \\
		Mode             & HBA\_DUAL\_INNER                       \\
		Pointing         &    eFEDS\_128, eFEDS\_131, \\
		 &   eFEDS\_134, eFEDS\_136, \\
		 &   eFEDS\_139, eFEDS\_142                              \\
	     &   P129+02, P132+02, \\
	     &   P134+02, P137+02, \\
	     &   P139+02, P126+02                         \\
	     &   G09\_A, G09\_B, \\
	     &   G09\_C, G09\_D   \\
		Calibrator              &      3C~196, 3C~295      \\
		Frequency (usable, MHz)            &  $120$--$168$       \\
		Central frequency (MHz)    &       144 \\
		Number of subbands (SB)    &       241             \\
		Bandwidth per SB (kHz)     &      195.3            \\
		Channels per SB            &        16             \\
		On-source time (hr)        &        184$^a$          \\  
		Integration time (s)       &        1                   \\
		Frequency resolution (kHz) &       12.2               \\
		Correlations               &  XX, XY, YX, YY             \\
		Number of stations         &        73--75 (48 split core, \\
		                           & 14 remote, 9--13 international$^{b}$)                     \\ 
		\hline\hline
	\end{tabular} \\
	Notes: $^a$: calculated from the total duration on all pointings, including simultaneous observations with two LOFAR beam;  $^{b}$: International stations are not used in this study.
	\label{tab:obs}
\end{table}

\begin{figure}[t!]
	\centering
	\includegraphics[width=1\columnwidth, height=0.7\columnwidth]{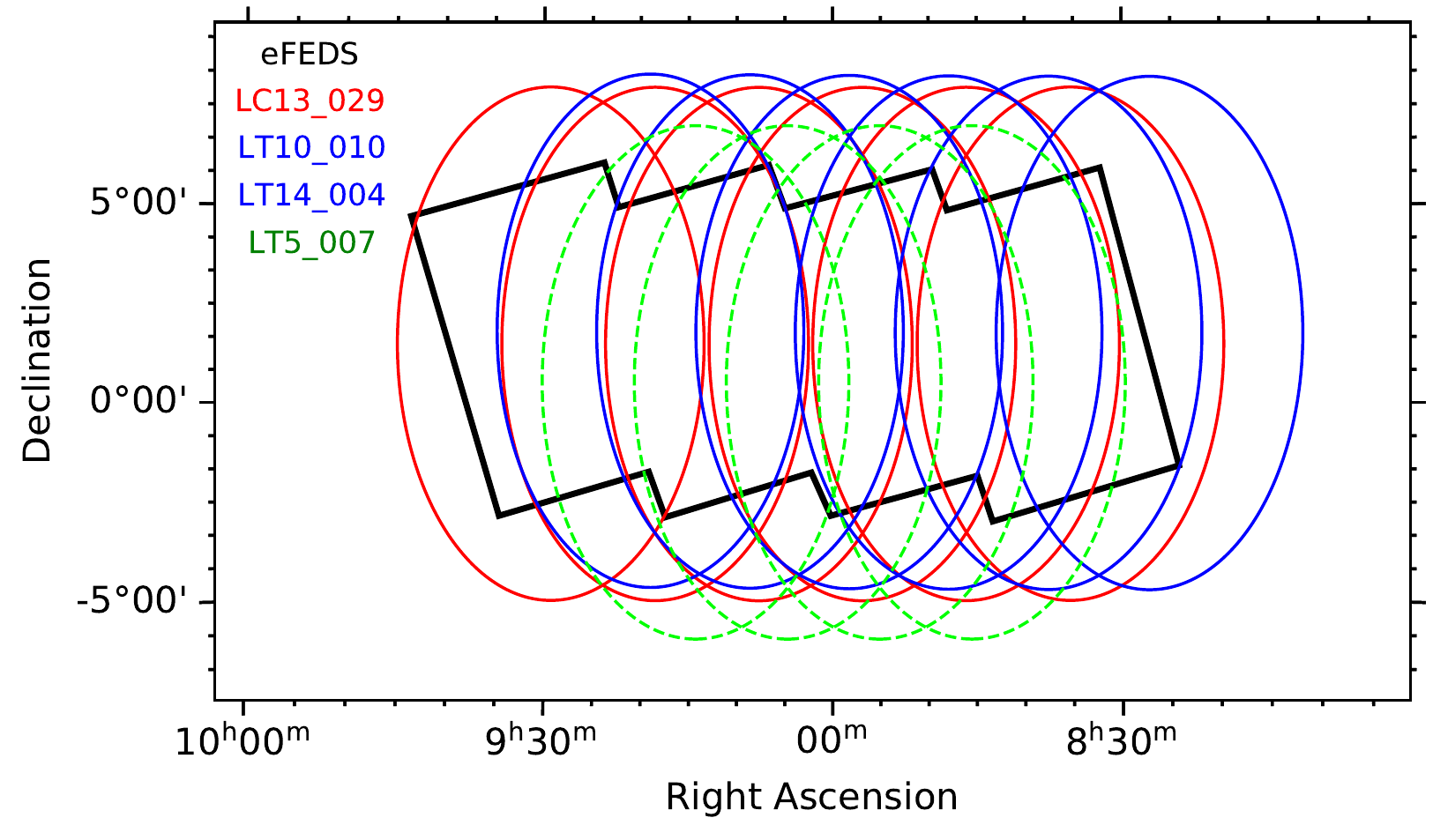}
	\caption{LOFAR observations of the eFEDS field, shown in the black region. The elliptical lines show the LOFAR pointing locations in the projects LC13\_029 (red), LT5\_007 (green), LT10\_010 (blue), and LT14\_004 (blue). The major and minor axes of the ellipse (i.e. 4.0~degree and 6.7~degree) are the FWHM of the LOFAR station beam along the RA and Dec axes, respectively.}
	\label{fig:layout}
\end{figure}

The eFEDS field was observed with the LOFAR High Band Antennae (HBA) for a total of 184 hours (including simultaneous observations by two LOFAR pointings) between February 24, 2016 and May 27, 2020 by projects LC13\_029 (100 hours), LT5\_007 (32 hours), LT10\_010 (44 hours), and LT14\_004 (8 hours). The eFEDS field is entirely covered by six pointings of LC13\_029 that are separated by 2.7 degree in a row. The LT5\_007 observations that are centred on the GAMA~09 field cover the central region of the eFEDS field with four pointings separated by 2.4 degree. LT10\_010 and LT14\_004, as part of the LOFAR Two-meter Sky Survey \citep[LoTSS;][]{Shimwell_2017,Shimwell_2019}, are positioned on the LoTSS grid where pointings are typically 2.58 degree apart. We present a layout of the LOFAR observations of the eFEDS field in Fig. \ref{fig:layout}. The setup for all observations is described in detail in \cite{Shimwell_2017,Shimwell_2019}. The observing frequency is from 120~MHz to 187~MHz, but we remove the data above 168~MHz where the signal is highly contaminated by RFI. Each pointing was performed by multiple chunks of 2 or 4 hours when the field is at high elevation (i.e. an average elevation of 35~degree). Bright radio sources 3C~196 and/or 3C~295 were observed for 10 minutes each before and after the observations of the target fields and are used as primary calibrators. Details of the observations are given in Table \ref{tab:obs}.

\begin{figure*}[t!]
	\centering
	\includegraphics[width=0.49\textwidth, height=0.49\textwidth]{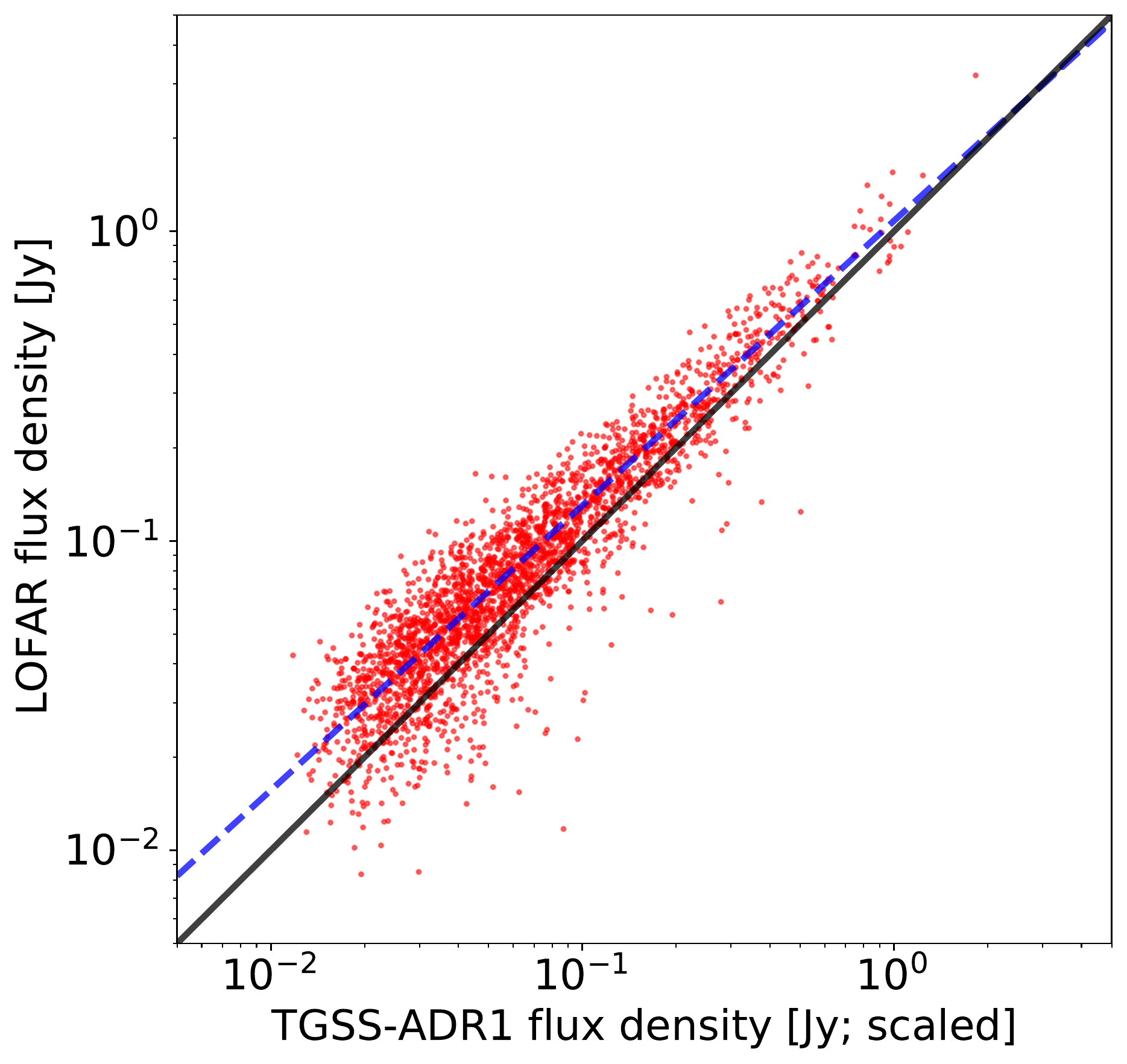} \hfill
	\includegraphics[width=0.49\textwidth, height=0.49\textwidth]{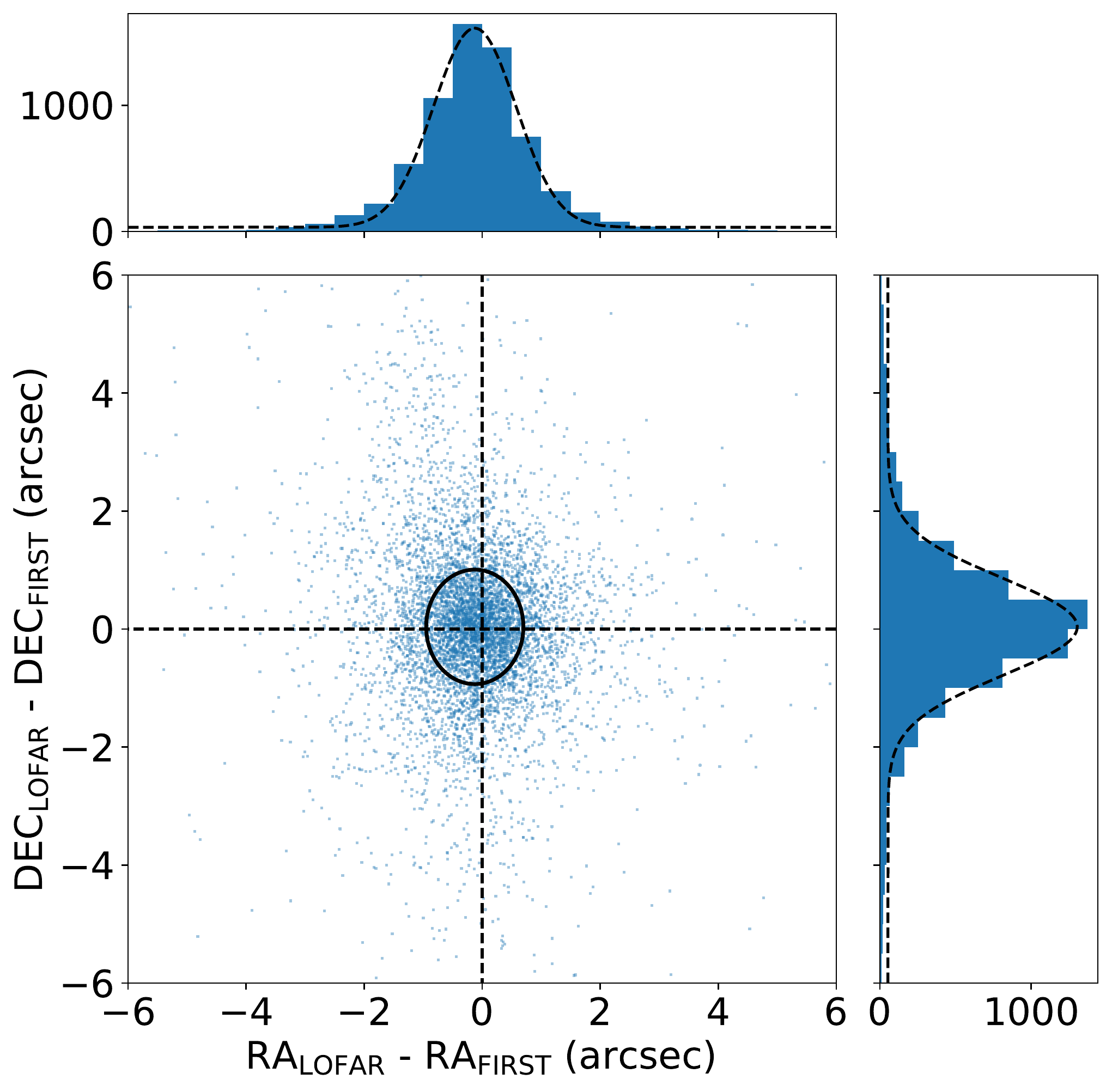}
	\caption{\textit{Left}: Scatter plot between the TGSS-ADR1 (scaled) and LOFAR flux densities for single-Gaussian sources. The dashed blue line is the best fit of the LOFAR and TGSS-ADR1 (scaled) flux densities, $\log_{10}{(S_{\rm LOFAR}})=0.92\times\log_{10}{(S_{\rm TGSS-ADR1;\,scaled}}) + 0.03$~[Jy]. The black solid line is a diagonal line with slope 1. \textit{Right}: RA and Dec offsets for the LOFAR and FIRST detected single-Gaussian sources. The histograms of the offsets, including the best-fit Gaussian dashed lines, are plotted in the top and right panels, respectively. The ellipse shows the peak location (i.e. $0.13\arcsec$ to the left and $0.04\arcsec$ to the top of the centre point) and the FWHM (i.e. $0.70\arcsec$ and $0.82\arcsec$ in RA and Dec.) of the Gaussian functions that are obtained from the fitting of the RA and Dec offset histograms.}
	\label{fig:fluxscale}
\end{figure*}

We adopt the standard calibration procedure that has been developed for LoTSS \citep[][]{Shimwell_2017,Shimwell_2019}. The calibration aims to correct for the direction-independent and direction-dependent effects (e.g. ionosphere and beam model errors) which need to be corrected for high-fidelty imaging with the LOFAR HBA. The data for each pointing were separately processed with \texttt{PREFACTOR}\footnote{\url{https://github.com/lofar-astron/prefactor}} \citep{VanWeeren2016a,Williams2016a,deGasperin2019} and \texttt{DDF-pipeline}\footnote{\url{https://github.com/mhardcastle/ddf-pipeline}} \citep{Tasse2014a,Tasse2014b,Smirnov2015,Tasse2018,Tasse2021}. In detail, the processing was identical to that described by \cite{Tasse2021} with one exception: in order to deal with the effect of sources outside the target $8 \times 8$ degree field but still covered by the very N-S elongated LOFAR primary beam, the first step of the pipeline for each image was to make a very large ($27 \times 27$ degree) image of the whole primary beam and subtract off sources detected by \texttt{DDFacet} that appeared in this image but lay outside the target field.

The pipeline produces high-resolution ($<9\arcsec$) images for each pointing with an rms noise of $\approx$170~$\mu$Jy~beam$^{-1}$ in the pointing centre and $\approx$335~$\mu$Jy~beam$^{-1}$ in the regions 2.5~degree from the pointing centres. Given the large LOFAR station beam (i.e. FWHM of 4~degree in an E-W direction at the central frequency of 144~MHz), the separation of 2.4--2.7~degree between the pointings leads to a significant overlap between the images. To increase the fidelity of the images, we convolved the images to a common resolution of $8\arcsec\times9\arcsec$ and made a mosaic of the entire eFEDS field in the manner described by \cite{Shimwell_2019}, reprojecting each image onto a $50,000 \times 27,000$ pixel image with 1.5-arcsec pixels centred on RA=9h, Dec=1 degree and then combining the reprojected images weighting by the local image noise at each pixel, taking account of the primary station beam. No astrometric blanking was carried out in the mosaicing and each image was corrected before mosaicing to the flux scale of \cite{roger:73} in the manner described by \cite{hardcastle:21}. The noise in the resulting mosaic is non-uniform but reduces to $\approx$135$\,{\rm \mu Jy~beam^{-1}}$ in the central parts of the image.

To produce a catalog of radio sources, we performed source detection on the high-resolution ($8\arcsec\times9\arcsec$) mosaic of the eFEDS field with the Python Blob Detector and Source Finder (\texttt{PyBDSF}\footnote{\url{https://github.com/lofar-astron/PyBDSF}};
\citealt{2015ascl.soft02007M}). Sources were detected with a peak detection threshold of $5\sigma$ (\texttt{thresh$\_$pix}=5) and an island threshold of $4\sigma$ (\texttt{thresh$\_$isl}=4) that limits the boundary for the source fitting. Here the local noise rms, $\sigma$, is calculated using a box of
($150\times150$) pixels$^2$ that slides across the mosaic with a step of $15\,{\rm pixels}$. Around bright sources, typically compact, where the pixel values are higher than $150\sigma$ ($\texttt{adaptive\_thresh}=150$), we used a smaller box of $(60\times60) \ {\rm pixels}^2$ and a sliding step of $15 \ {\rm pixels}$. The smaller box is more accurate for the estimate of the high noise rms around bright sources. The source detection produces a catalog of 45,207 sources, most of which (99.6 percent) have 144~MHz flux densities below $1\ {\rm Jy}$. 

The mosaic that is made with the standard procedure described above typically has a flux density uncertainty of 10 percent. However, to further check the flux scale in the eFEDS mosaic we compare the integrated flux densities of the LOFAR detected sources with those in the TGSS-ADR1 (TIFR GMRT Sky Survey - Alternative Data Release 1, \citealt{Intema_2017}) 150~MHz data, which has similar central frequency. The LOFAR mosaic was smoothed to the resolution of the TGSS-ADR1 (i.e. $25\arcsec$) and regridded to match the spatial dimensions of the TGSS-ADR1 image. Radio sources in the LOFAR and TGSS-ADR1 $25\arcsec$ images are detected with \texttt{PyBDSF} in an identical manner as done for the LOFAR $8\arcsec\times9\arcsec$ mosaic above. There are 4,585 sources detected with both LOFAR and TGSS-ADR1 observations. Sixty percent of these sources (i.e. 2,695) are modeled by a single Gaussian and are used for the flux scale comparison. Since the observing frequencies for the LOFAR and TGSS-ADR1 data are different, we rescaled the flux densities of the TGSS-ADR1 sources to match those at the frequency of the LOFAR data (144~MHz) by assuming a common spectral index of $0.8$ (see Sec. \ref{sec:construction} for a definition). We performed a linear fit to the LOFAR and TGSS-ADR1 scaled flux densities, weighting by the LOFAR flux densities, and obtained a relation $\log_{10}{(S_{\rm LOFAR}})=0.92\times\log_{10}{(S_{\rm TGSS-ADR1;\,scaled}}) + 0.03$~[Jy]. The integrated flux densities of the radio sources in the LOFAR catalog is $\sim$10 percent higher than those in the TGSS-ADR1 catalog. In this paper, we assume an uncertainty of 20 percent for the integrated flux densities of the LOFAR detected sources. In Fig. \ref{fig:fluxscale}, we present a scatter plot between the flux densities of the LOFAR and TGSS-ADR1 detected sources. The LOFAR detected sources, especially the faint ones, have higher flux densities than those found with the TGSS-ADR1 observations. 

Following \cite{Shimwell_2019}, we checked the astrometry of the sources detected with  \texttt{PyBDSF} in the LOFAR $8\arcsec\times9\arcsec$ mosaic by comparing their locations with those of their FIRST~1.4~GHz counterparts. We use the FIRST survey due to its high astrometric accuracy of $0.1\arcsec$ compared to the absolute radio reference frame \citep{White1997} and the comparable spatial resolution of both surveys (i.e. $5\arcsec\times5\arcsec$ for FIRST and $8\arcsec\times9\arcsec$ for LOFAR). We cross-matched the sources within a radius of $9\arcsec$ in the LOFAR and FIRST catalogs and found 10,709 common sources, of which 6,601 are single-Gaussian LOFAR sources. We calculate the offsets in RA and Dec for these single-Gaussian sources and present them in Fig. \ref{fig:fluxscale}. The histograms of the RA and Dec offsets are fitted with a Gaussian whose location and standard deviation are defined as the systematic offsets and total astrometric uncertainty. There are systematic offsets of $0.13\arcsec$ and $0.04\arcsec$ in RA and Dec, respectively. The standard deviations of the offsets in RA and Dec are $0.70\arcsec$ and $0.82\arcsec$, respectively. When comparing to the offsets between FIRST and LoTSS sources \citep{Shimwell_2019}, our results on the RA and Dec offsets are a factor of two to seven higher, and the standard deviations are a factor of two to three higher. These are likely due to the lower declination of the eFEDS field, as compared with the declination of $\approx50^\circ$ of the LoTSS-DR1 field, that results in a larger elongated beam and slightly more disturbed ionospheric conditions. However, the uncertainties are well within the resolution of the LOFAR observations (i.e. $8\arcsec\times9\arcsec$).


\subsection{Sample construction and properties}
\label{sec:construction}

The catalog of radio sources was cross-matched with the BCG positions (see Sec. \ref{sec:efeds}) by setting a sky threshold $\sim 3\theta$, with $\theta$ being the synthesised beam of the interferometric radio observation. The results were then manually inspected to check for the presence of false positives (i.e. radio sources incorrectly associated with an optical BCG) or false negatives (i.e. radio emission lying at more than 3$\theta$ from the BCG, but with an obvious association to it). We find no wrong BCG-radio association, while two clusters were initially mistakenly classified as non-detections.
To limit contamination, we applied the same cut $f_{\mathrm{cont},i} < 0.3$ discussed in Sec. \ref{sec:efeds}. According to Eq.~\ref{eq:fcont}, this implies that we are statistically allowing for 6\% contamination. This value, albeit conservative, produces a relatively small impact on our results.

The final catalog contains a total of 227 clusters, with only $\sim$1\% (3 out of 230) of objects lost to contamination. This is consistent with our expectations, since the cut we applied should result in a cluster catalog that is $\sim$99\% complete (see \citealt{Klein_2021} and \citealt{LiuAng_2021}). Out of the parent sample of 542 X-ray clusters, 312 did not match any of LOFAR radio sources. After applying the same contamination criteria, we were left with 248 clusters with no radio emission detected, losing $\sim$21\% of the original sample. These were then treated as radio upper limits assuming a flux limit of 3$\sigma$, where $\sigma$ is the local $rms$ noise of the LOFAR mosaic at the position of the cluster. The increase in the number of clusters lost to contamination with respect to detections is easy to explain, once it is considered that excluded objects are not real clusters, but mostly contaminants (e.g. bright AGN). Therefore, it is less likely to find a radio counterpart. Again, we refer to \citet{Klein_2021} and \citet{LiuAng_2021} for further details.

Nevertheless, not every cluster/group, in reality, hosts radio galaxies. In fact, some groups only contain a few ($< 10$) galaxies, and only $\sim$1\% of all observed galaxies are active \citep{Padovani_2017}. This fraction should also be significantly higher in overdense environments such as clusters. \citet{Sabater_2019} found that 100\% of their sample of AGN in massive galaxies ($> 10^{11}$ M$_\odot$) are always switched on above a 144 MHz luminosity of $10^{21}$ W Hz$^{-1}$. In fact, it has been observed that there is a strong link between the presence of radio AGN activity and the host galaxy mass \citep{Best_2005b, Sabater_2013}. As already discussed in Sec. \ref{sec:intro}, \citet{Kolokythas_2018, Kolokythas_2019} report rates at 235 MHz of 92\% and 82\% for their sample of 26 and 27 galaxy groups, respectively. P20 report a detection rate for COSMOS groups of $\sim$70\%, with $rms \sim 12\mu$Jy beam$^{-1}$. Here, the same fraction is only 48\% (given the cut we applied for contamination). This is likely due to the lower Signal-to-Noise ratio (S/N) of LOFAR eFEDS with respect to the single-target observations used to build CLoGS, while P20 exploited the VLA-COSMOS Deep Survey. Furthermore, CLoGS was built with low-redshift ($z < 0.02$) groups, while our sample reaches $z \sim 1.3$.

The luminosity of all the radio sources, including upper limits, was estimated as:
\begin{equation}
    L_{144\text{MHz}} = S_{\text{144}} 4\pi D_{\text{L}}^2  (1+z)^{\alpha-1} ,
    \label{eq:radiolum}
\end{equation}
where S$_{\text{144}}$ is the flux density at 144 MHz, D$_{\text{L}}$ is the luminosity distance at redshift $z$ and $\alpha$ is the spectral index $S_{\nu} \propto \nu^{-\alpha}$, assumed $\sim$ 0.8 for all radio galaxies since we are at low frequency and most sources show a relatively extended morphology, rather than being compact and point-like as usually observed at higher frequency.

\begin{figure*}[t!]
	\centering
	\includegraphics[height=28em, width=26em]{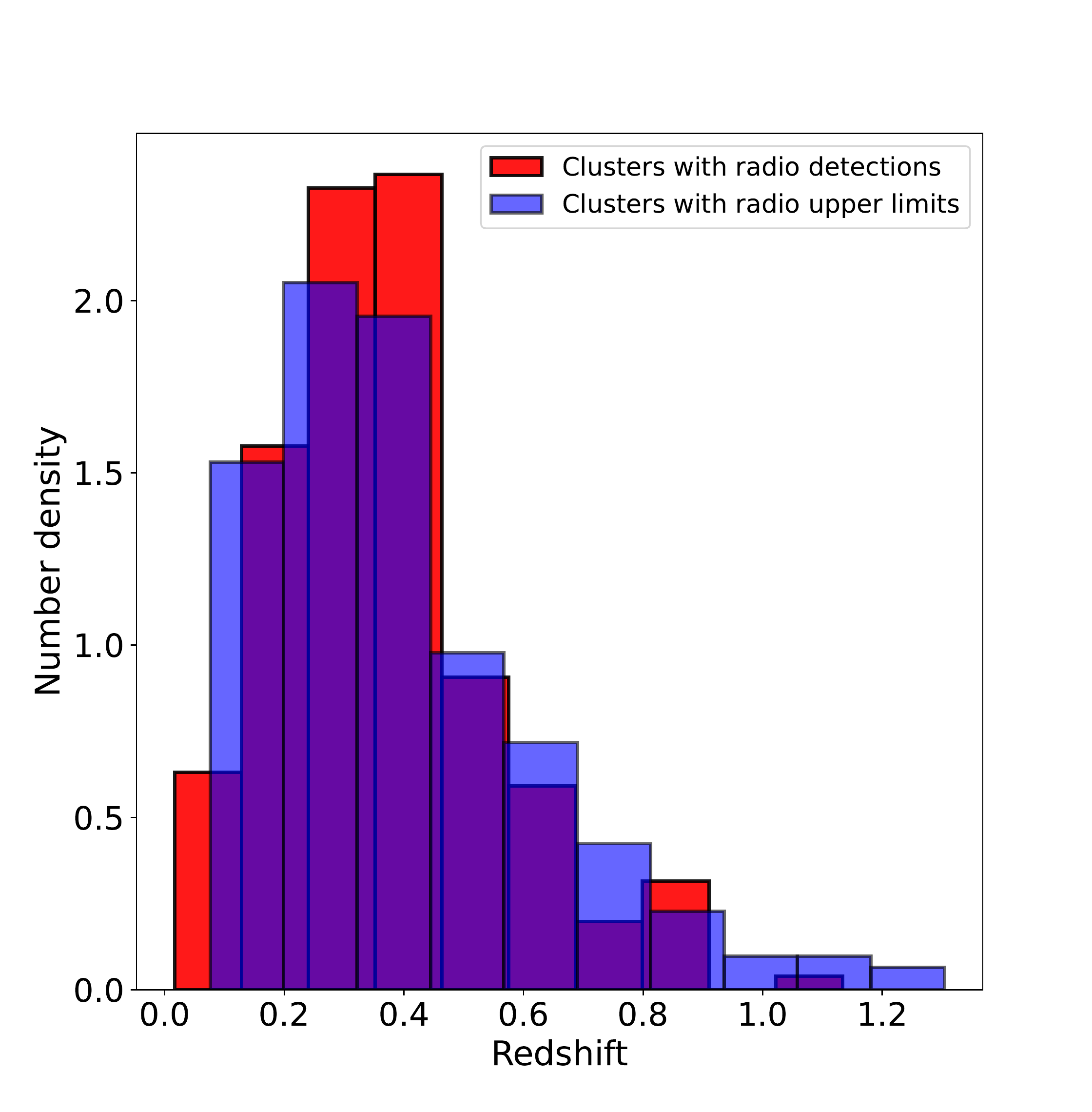}
	\includegraphics[height=28em, width=26em]{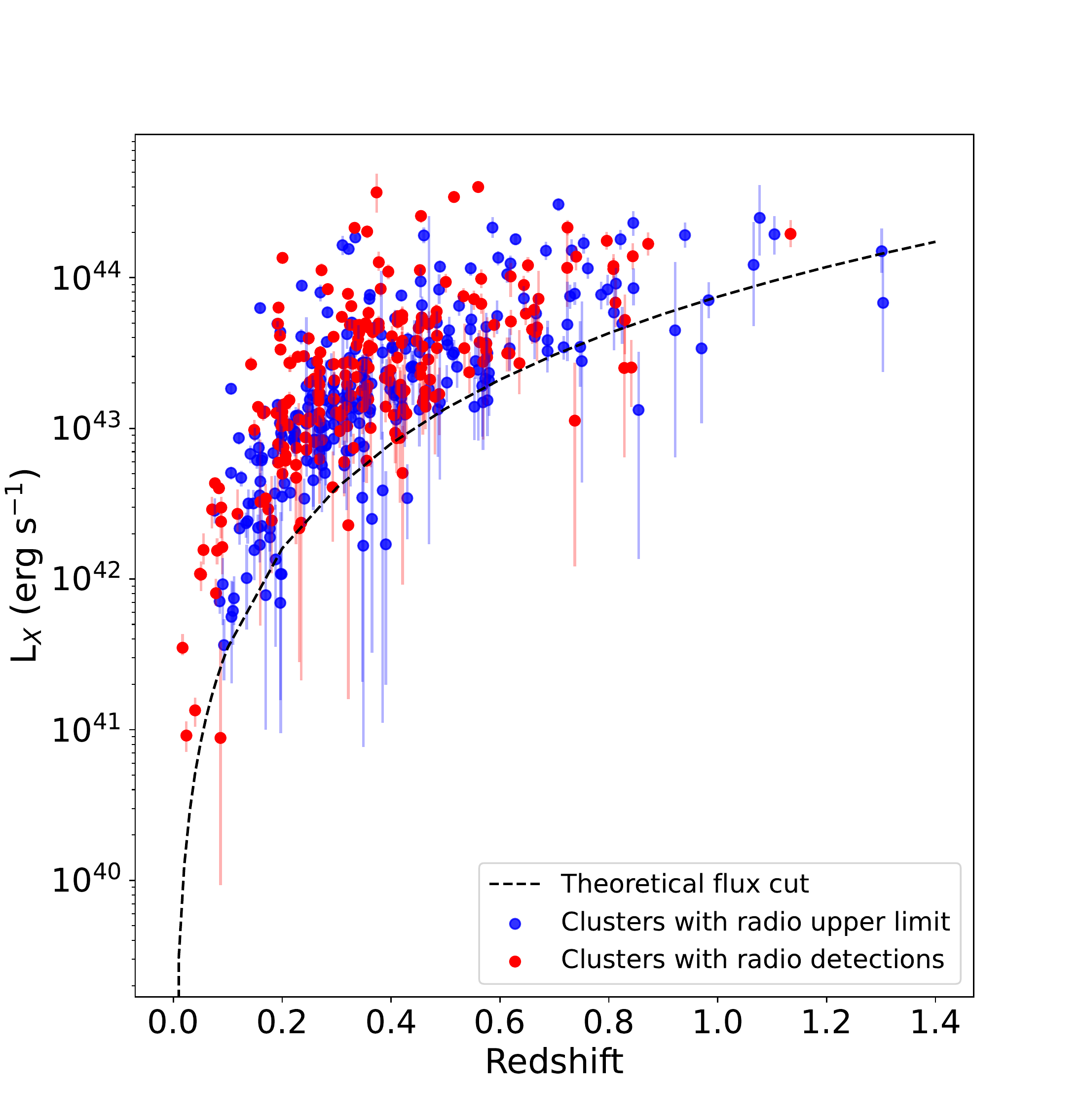}
	\caption{\textit{Left}: Histogram showing the redshift distribution of the sample, classified into radio detections (red) and upper limits (blue). \textit{Right}: $L_{X, 500}$ vs. redshift for the sample. The dashed line denotes the theoretical flux cut of the eROSITA observation.}
	\label{fig:reddistr}
\end{figure*}

The left panel of Fig.~\ref{fig:reddistr} presents the redshift distribution for the sample, classified into detections and radio upper limits. The detection and non-detection distributions look similar up to $z \sim$ 0.9. The highest-$z$ detection is at $z \sim$ 1.1, while there is one radio upper limit at $z \sim$ 1.3. The right panel shows $L_{X, 500{\rm kpc}}$ vs. redshift, with the same classification, with $L_{X, 500{\rm kpc}}$ being the 0.5-2.0 keV luminosity measured within a 500 kpc radius. The flux sensitivity is $F_X = 1.5 \times 10^{-14}$ erg s$^{-1}$ cm$^{-2}$. Further details on the eROSITA selection function and completeness can be found in \citet{LiuAng_2021}.


\section{Analysis and discussion}
\label{sec:analysis}

\subsection{X-ray and radio luminosity distributions}

\begin{figure*}[!]
	\centering
	\includegraphics[height=28em, width=26em]{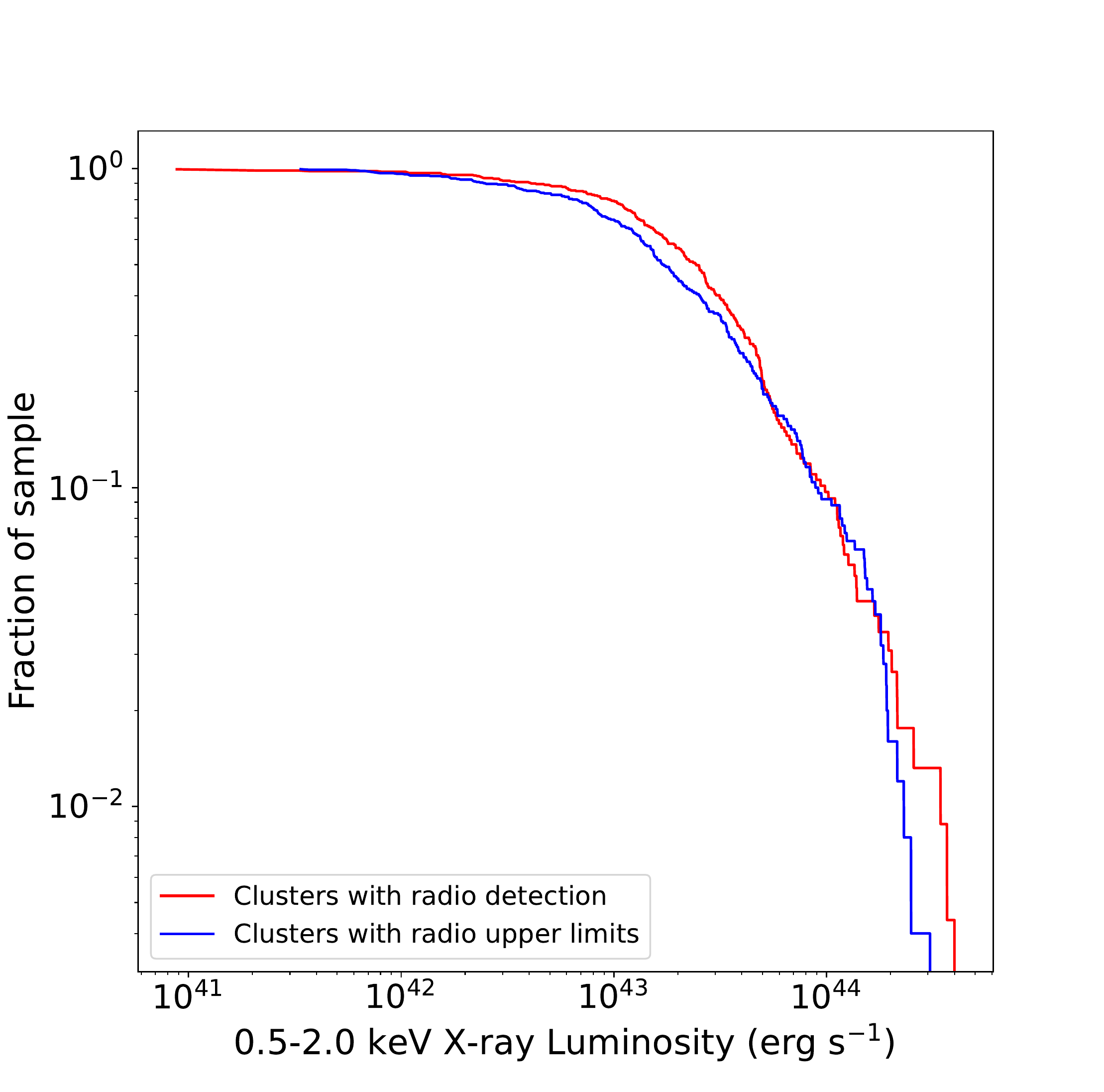}
	\includegraphics[height=28em, width=26em]{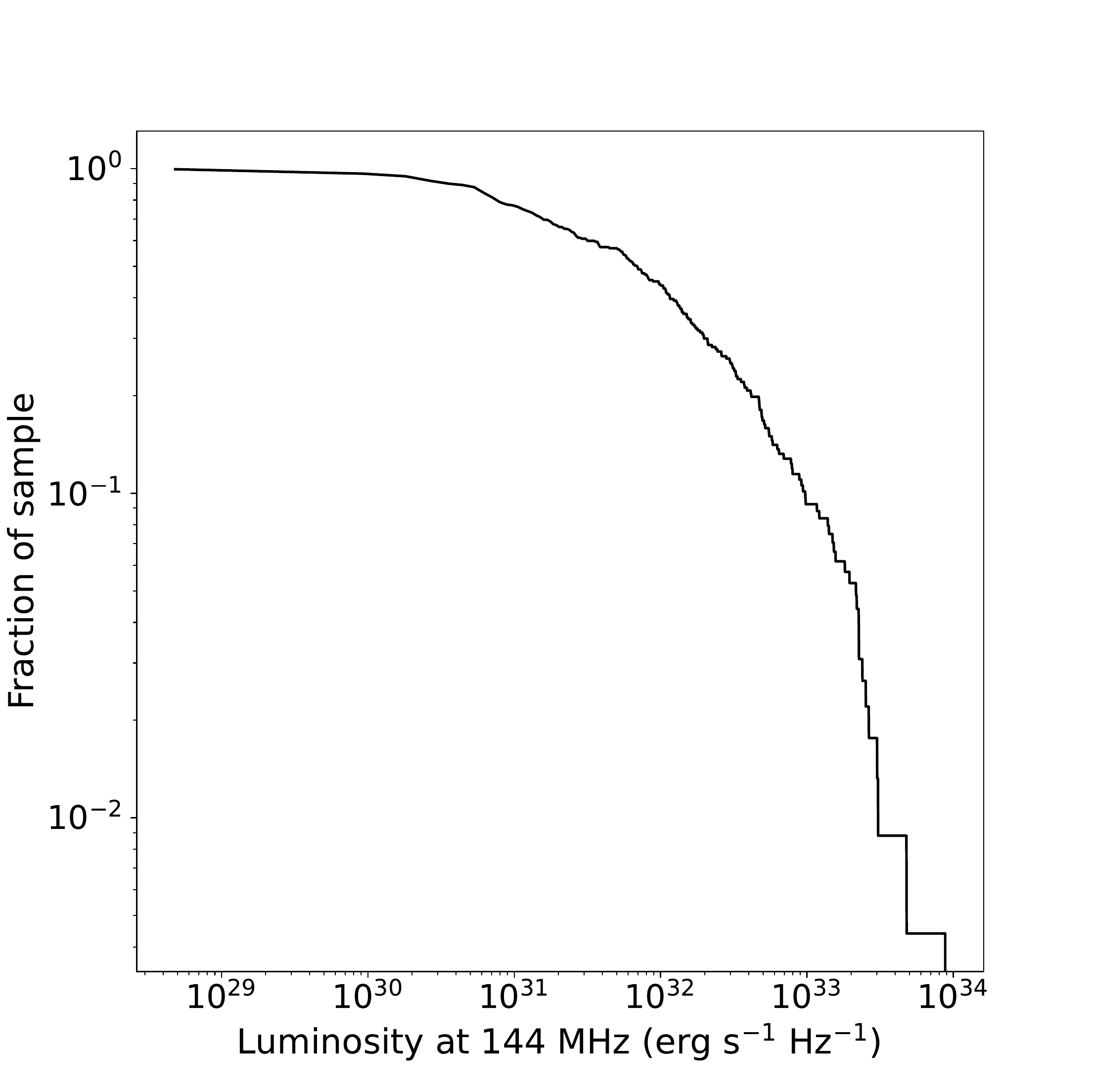}
	\caption{\textit{Left}: X-ray luminosity distribution for the parent sample of clusters and groups, divided into objects with (red) and without (blue) radio detection. \textit{Right}: Radio luminosity distribution for clusters and groups with radio detection.}
	\label{fig:lumfunc}
\end{figure*}

In Fig.~\ref{fig:lumfunc} we show the X-ray and radio luminosity distributions. The X-ray distribution, in the left panel, spans the range from $L_{X, 500{\rm kpc}} \sim$ 10$^{41}$ erg s$^{-1}$ to $4 \times 10^{44}$ erg s$^{-1}$ for objects with radio detections, while the range for clusters with upper limits is slightly narrower, reaching $3 \times 10^{44}$ erg s$^{-1}$. Due to the high sensitivity of eROSITA, we are able to reach lower luminosities than the existing X-ray samples of clusters and groups. The BCS sample, compiled with \textit{ROSAT} \citep{Ebeling_1997}, reaches $L_X \sim 10^{42}$ erg s$^{-1}$, similarly to the REFLEX II catalog \citep{Bohringer_2014}. On the other hand, our upper range is lower than both the BCS and the REFLEX II, which go well beyond $L_X \sim 10^{45}$ erg s$^{-1}$, since our sample comes from a relatively small field in the sky. The forthcoming eROSITA all-sky survey (eRASS, Bulbul et al. in prep.) will observe a large number of clusters and groups, allowing to extend our analysis to higher luminosities.

The radio luminosity distribution at 144 MHz, in the right panel, ranges from $L_{\rm 144MHz} \sim 10^{29}$  erg s$^{-1}$ Hz$^{-1}$ to $\sim 10^{34}$ erg s$^{-1}$ Hz$^{-1}$. 
Given the assumption on the spectral index made above, the upper range of luminosities at 144 MHz corresponds to $L_{\rm 1.4GHz} \sim 1.6 \times 10^{33}$ erg s$^{-1}$ Hz$^{-1}$. This is lower than other samples that have recently been studied at this frequency. The catalog of 1.4 GHz radio sources in  galaxy groups analysed in P20 reaches $L_{\rm 1.4GHz} \sim 10^{34}$ erg s$^{-1}$ Hz$^{-1}$, similarly to the sample of BCG radio galaxies by \citet{Hogan_2015}. Finally, we note that the sample studied at 235 MHz by \citet{Kolokythas_2018, Kolokythas_2019} ranges from $L_{\rm 235MHz} \sim 10^{27}$ to $10^{32}$ erg s$^{-1}$ Hz$^{-1}$. Converting from 144 MHz luminosity, eFEDS radio galaxies span from $L_{\rm 235MHz} \sim 6.8 \times 10^{28}$ to $6.7 \times 10^{33}$ erg s$^{-1}$ Hz$^{-1}$. Therefore, our sample extends to higher radio powers, but does not go as deep as CLoGS. Nevertheless, it consists of 227 clusters and groups, compared to the 53 groups that belong to CLoGS.


\subsection{BCG offsets}

\begin{figure}
	\centering
	\includegraphics[height=28em, width=26em]{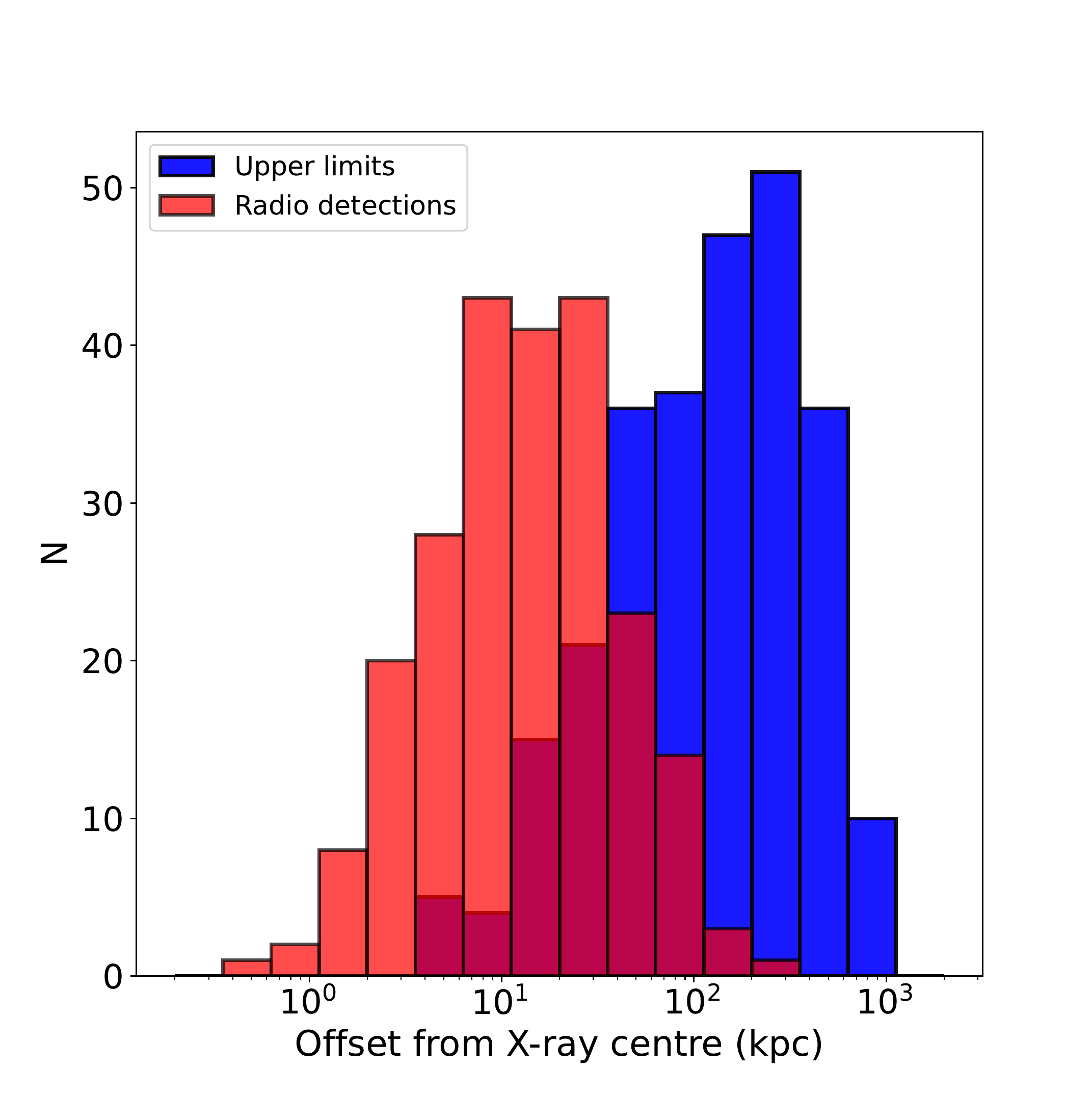}
	\caption{Histogram showing BCG offsets from the X-ray emission peak, assumed as the centre of the cluster/group, for galaxies with AGN radio emission (red) and for those with radio upper limits (blue).}
	\label{fig:offset}
\end{figure}

Fig.~\ref{fig:offset} shows the histogram of the BCG offset from the centre of the host cluster/group. The centre was estimated by fitting a two-dimensional $\beta$-model \citep{Cavaliere-Fusco_1976} to the X-ray emission. Most BCGs with detected AGN radio emission lie within $\sim$50 kpc from the cluster centre ($\sim$84\%). For these clusters, the median value of the offset distribution is $\sim$15 kpc, with dispersion $\sim$30 kpc. At larger offsets it is easier to find BCGs that do not host a radio galaxy. For clusters with no radio detection, the median is $\sim$130 kpc with dispersion $\sim$190 kpc.

Small offsets ($<$50 kpc) are expected and found in most relaxed clusters since even a minor merger can induce sloshing and displace the X-ray emission peak from the BCG \citep[e.g.,][]{Hamer_2016, Pasini_2019, Ubertosi_2021, Pasini_2021}. Large offsets (100-1000 kpc) are often an indication of a strongly disturbed cluster environment \citep[][and references therein]{Rossetti_2016, dePropris_2021}. The relation between BCGs, the triggering of the AGN and the offset from the cluster centre has been widely discussed and was recently studied in \citet{Pasini_2021b}. In that paper, the authors found that it is more common for more central BCGs to show radio-loud AGN since in these galaxies the accretion onto the central BH is boosted by the strong cooling in the cluster core. Similar results have also been discussed in \citet{Burns_1990, Best_2007, Cavagnolo_2008, Shen_2017}. On the other hand, off-centre galaxies have to rely on more episodic processes, such as cluster/group mergers and/or galaxy interactions. We find the same results in this sample since, as discussed above, radio-loud AGNs are mostly found at offset $<$50 kpc.


\subsection{The extent of BCG radio galaxies}

\begin{figure}
	\centering
	\includegraphics[height=28em, width=26em]{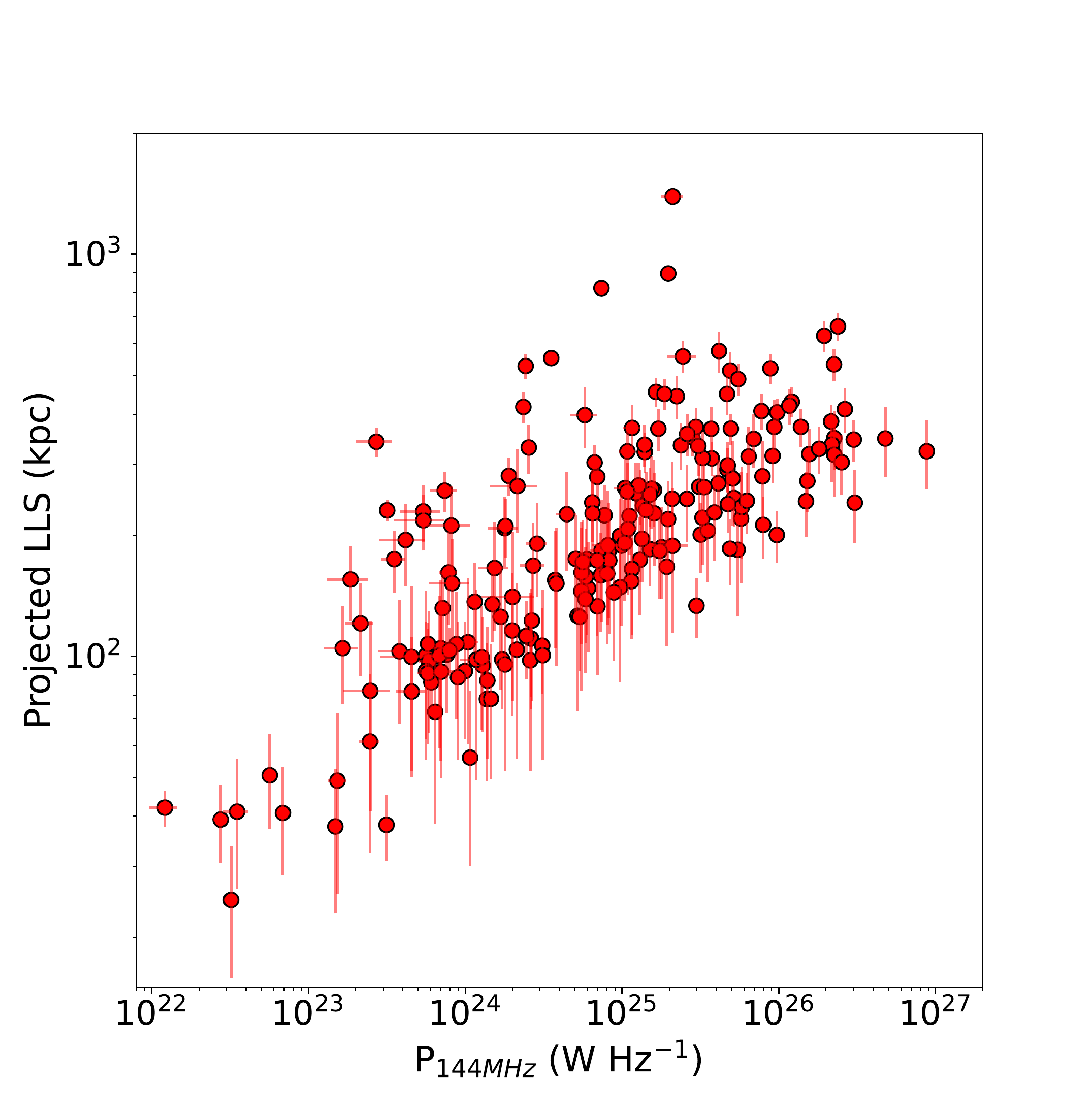}
	\includegraphics[height=28em, width=26em]{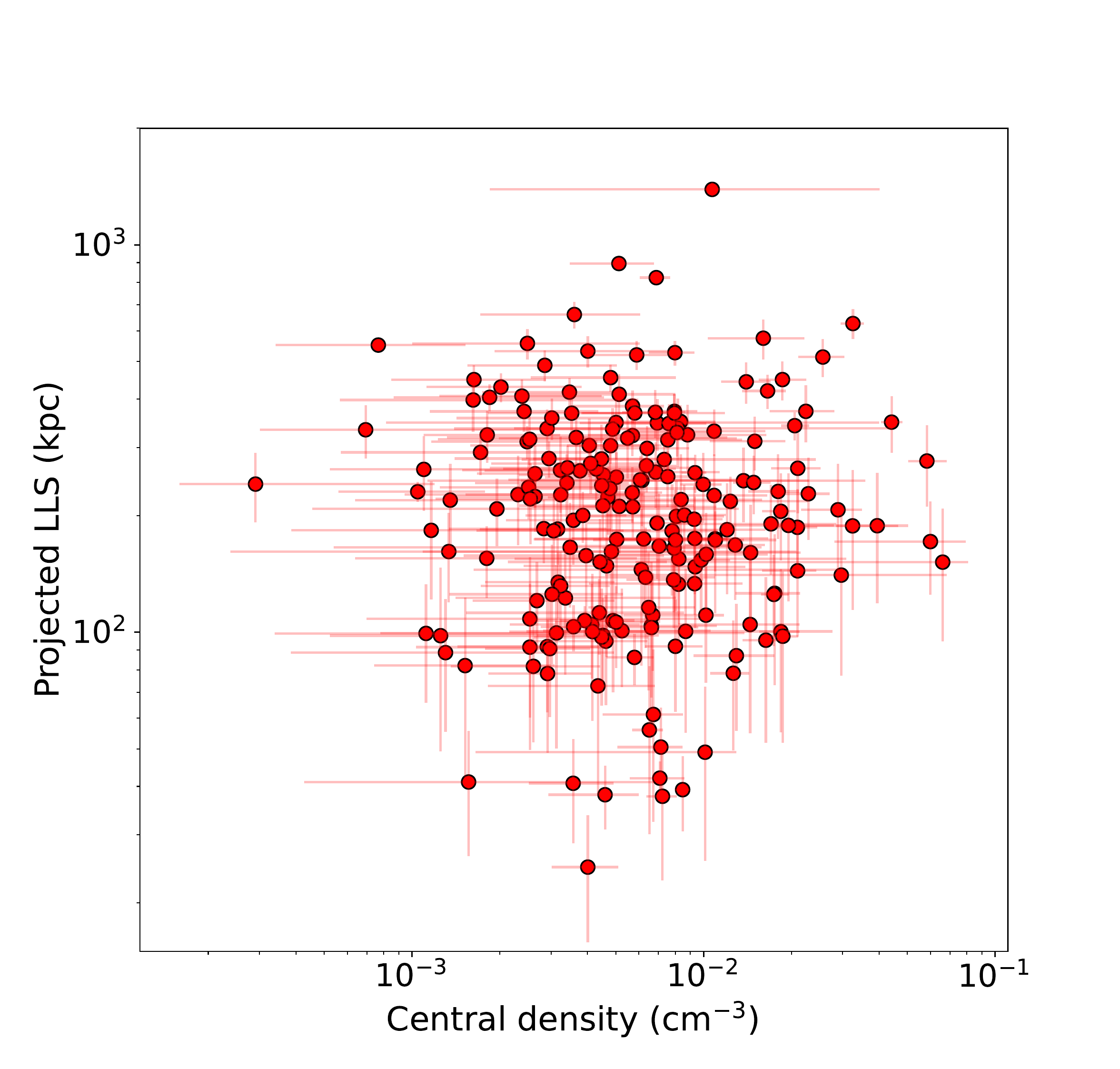}
	\caption{\textit{Top}: Projected Largest Linear Size (LLS) vs. radio power of eFEDS radio galaxies. \textit{Bottom}: Projected LLS of the radio galaxy vs. ICM central density.}
	\label{fig:growth}
\end{figure}

Radio galaxies exhibit a plethora of different shapes and sizes. The reasons for the unusual size of some giant radio galaxies \citep[e.g.,][]{2021A&A...647A...3B, Dabhade_2020} and for the significantly smaller extent of some others (e.g. FR0, \citealt{Baldi_2015}) have been investigated previously. \citet{Hardcastle_2019} presented the largest up-to-date sample of radio galaxies in which the relation between the radio power and the linear size was investigated. The location of a source in this diagram is indicative of its initial conditions and current evolutionary state \citep{Hardcastle2018}. \citet{Kolokythas_2018} found a clear link between the 235 MHz power and the projected Largest Linear Size (LLS) of their resolved radio galaxies. The same relation was already found for cluster and field radio galaxies by \citet{Ledlow_2002}, and is investigated for our sample of BCGs at 144 MHz (top panel of Fig.~\ref{fig:growth}). The LLS of radio galaxies was manually measured from the LOFAR eFEDS mosaic, assuming error equal to the synthesised beam. To exclude unresolved sources, only those with Largest Angular Size LAS $> 2\times$beam are taken into account.

Most sources show LLS between 100 and 300 kpc, with the mean at LLS$\sim$235 kpc and standard deviation $\sim$ 160 kpc. Large sources mostly show a classical double-lobed morphology, while the smallest ones are point-like. As previously observed, there is a positive correlation between LLS and luminosity, with larger radio galaxies being more powerful. We see that the relation holds even at relatively high luminosities\footnote{CLoGS only reaches $L_R \sim 10^{25}$ W Hz$^{-1}$ at 235 MHz.}. Nevertheless, we note that we are likely missing large, low-power radio sources because of surface brightness limitations. This issue has been extensively addressed in \citet{Hardcastle_2019} making use of a significantly larger sample (23344 objects) of radio galaxies.

Multiple environmental factors are likely to contribute to the size of the radio source. The most important one is the age, which necessarily introduces scatter into any relation with other physical quantities. Other factors include the location of the galaxy within the host cluster, the density of the ICM at the position of the galaxy, the efficiency of the accretion onto the AGN, the radio power of the outburst and others \citep[see e.g.][and references therein]{Moravec_2020}.

To this end, in the bottom panel of Fig.~\ref{fig:growth} we show the LLS of the radio galaxy plotted against the central density (at $R = 0.02R_{500}$) of the host cluster, obtained by fitting the cluster model by \citet{Vikhlinin_2006} to density profiles (see G21 for further details). We see no correlation of the LLS with the central density, suggesting that radio power is more prominent than ambient density in determining the size of the radio galaxy and that the contribution of other factors could affect a possible link. 


\subsection{Correlation between X-ray and 144 MHz radio luminosity}

In P20, we have studied the correlation between the 1.4 GHz power of radio galaxies and the X-ray luminosity of the host group for 247 galaxy groups in COSMOS. A similar correlation between the mass of galaxy clusters, known to correlate with the X-ray luminosity \citep[e.g.,][]{Lovisari_2020}, and the radio power of BCGs has been found by \cite{Hogan_2015}. Here, we focus on the same relation, albeit at the lower radio frequency of 144 MHz.

\begin{figure}
	\centering
	\includegraphics[height=28em, width=26em]{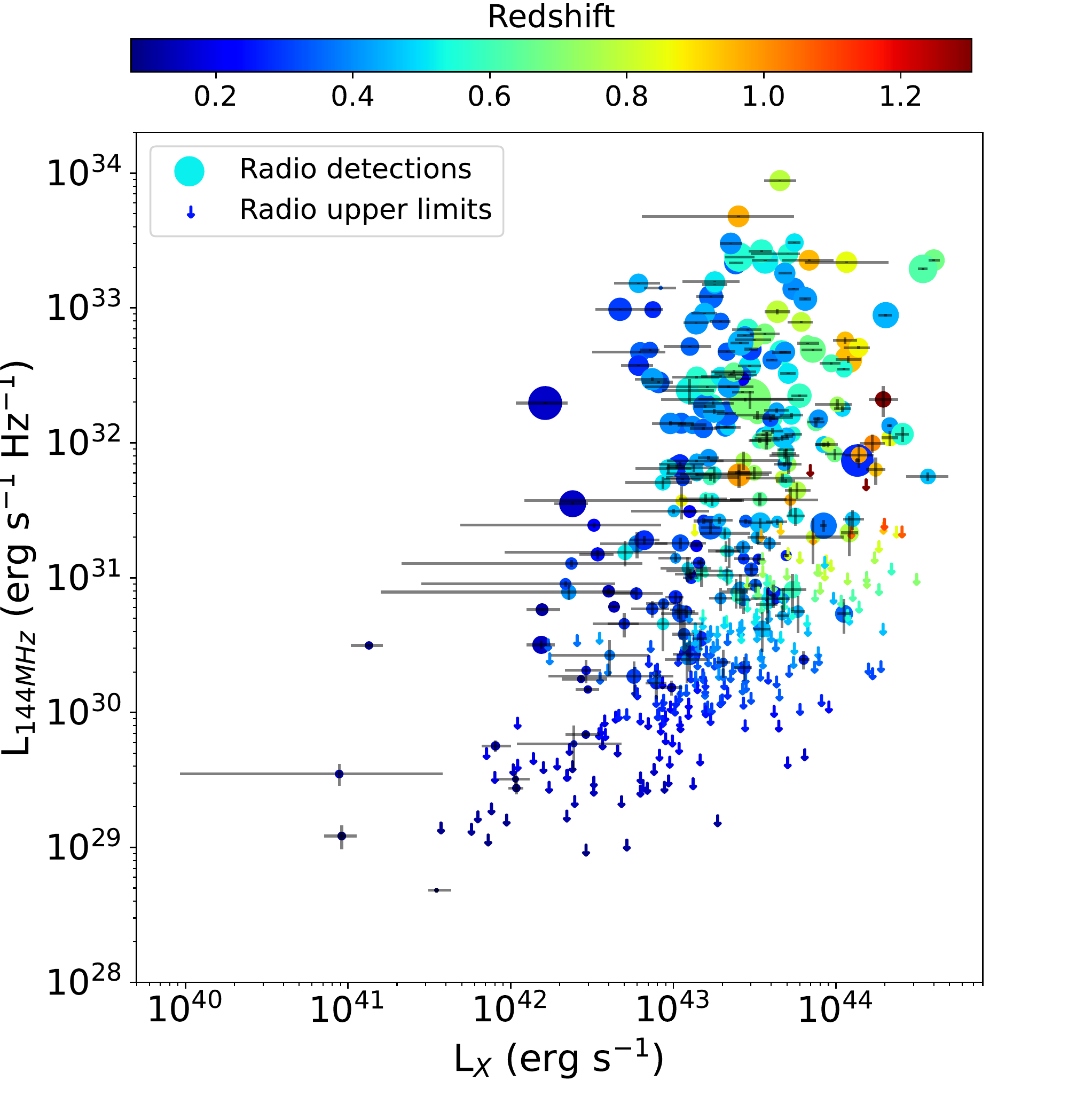}
	\caption{144 MHz power of radio galaxies vs. X-ray luminosity of the host cluster. Symbol sizes are proportional to the LLS of the source and their color indicates the redshift. Downward pointing arrows denote radio upper limits. Bars represent errors on both axes.}
	\label{fig:lxlr}
\end{figure}

Fig.~\ref{fig:lxlr} shows the 144 MHz power of the radio galaxy plotted against the X-ray luminosity of the host group/cluster. The size of the symbols is proportional to the LLS of the radio sources, and the colour corresponds to the redshift. Upper limits are represented by downward pointing arrows. There is a clear trend for stronger radio galaxies to be hosted in more X-ray luminous clusters, as found by P20. However, the significant number of radio upper limits makes it harder to determine whether the observed correlation is real or produced by selection effects set by the sensitivity of the observation.

To ascertain if the correlation is genuinely detected, we performed the partial correlation Kendall's $\tau$ \citep{Akritas_Siebert_1996} test. This tool has already been used in a number of papers \citep[e.g.,][]{Ineson_2015, Pasini_2020} to test correlations in the presence of upper limits and redshift-dependence. The algorithm estimates the null-hypothesis probability that selection effects are producing the correlation. If the probability is low, then it is likely that the correlation is real. The test performed on our sample gives a null-hypothesis probability $p < 0.0001 \%$ ($\tau = 0.1178$, $\sigma = 0.0227$), indicating that the correlation is real and not generated by selection effects. This result is consistent with P20, who also found that such a correlation, but at higher frequency, was not produced by biases. 

\begin{figure}
	\centering
	\includegraphics[height=28em, width=26em]{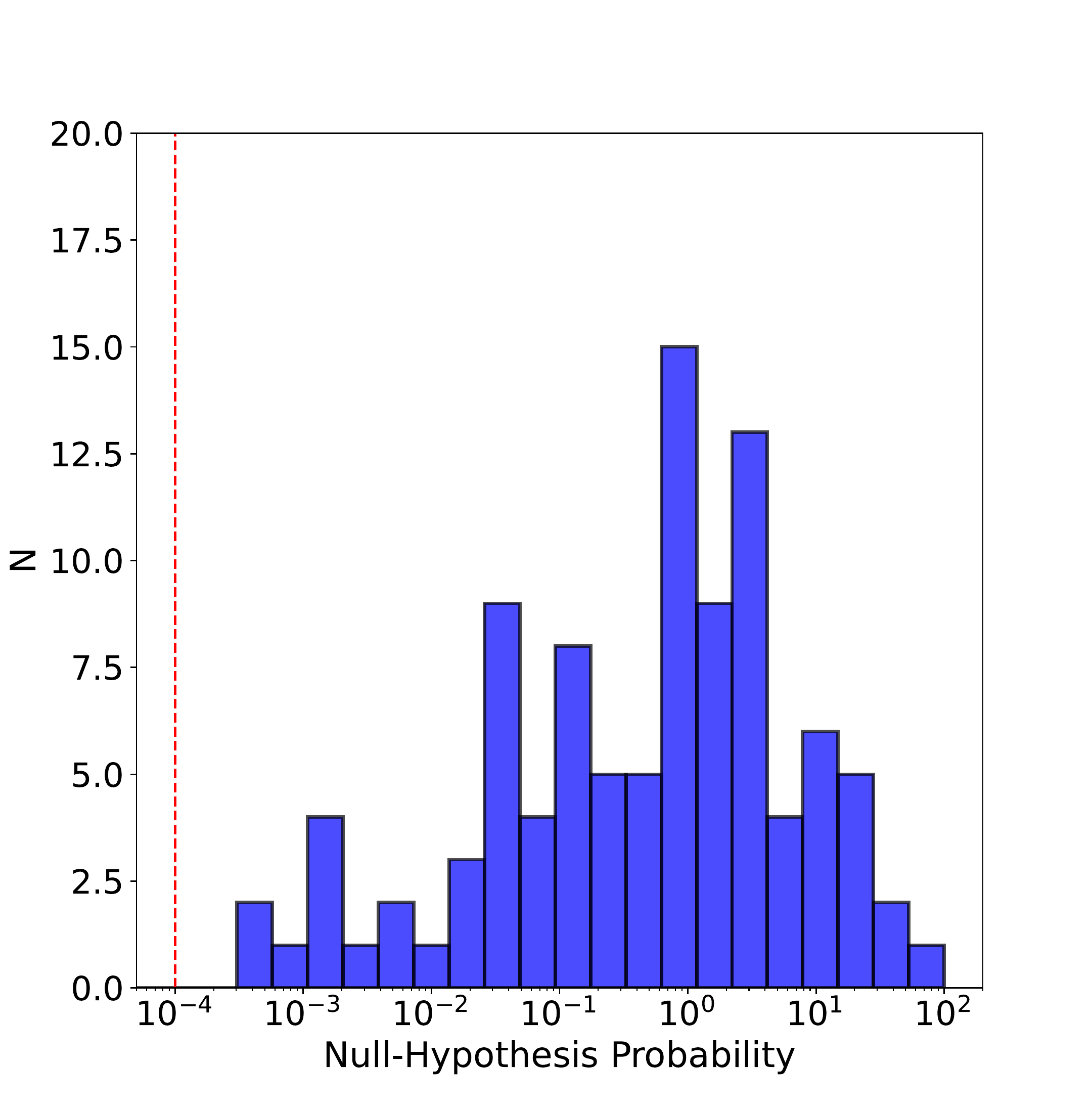}
	\caption{The histogram shows the null-hypothesis probability distribution that the correlation is not real for 100 mock datasets produced by the scrambling test. The red dashed line indicates the probability of the real sample.}
	\label{fig:pval}
\end{figure}

\citet{Bianchi_2009} argued that the Kendall's $\tau$ test may underestimate the redshift contribution, particularly when it comes to determining the significance and the functional relation. For this reason, they performed a `scrambling' test that has also been used in other works \citep[e.g.,][]{Bregman_2005, Merloni_2006}. The principle of this algorithm is to keep each $L_X/z$ pair since their association comes from the source selection. Then they shuffle the corresponding radio fluxes, assigning them to a new $L_X/z$ pair. The new radio luminosity is then computed at the new redshift (see Eq.~\ref{eq:radiolum}). If the correlation is real, one expects that it disappears when shuffling the luminosity pairs. We applied this test 100 times and for each cycle estimated the null-hypothesis probability through the Kendall $\tau$ test. Results are shown in Fig.~\ref{fig:pval}. Out of 100 cycles, the null-hypothesis probability is never found to be lower than the real sample. The mean probability value lies at $\sim$4\%, with a standard deviation of $\sim$9\%, while the peak lies between 0.7\% and 5\%. This result supports the hypothesis that the observed correlation is real.

\subsection{The X-ray/radio correlation at 1.4 GHz}

\begin{figure}
	\centering
	\includegraphics[height=28em, width=26em]{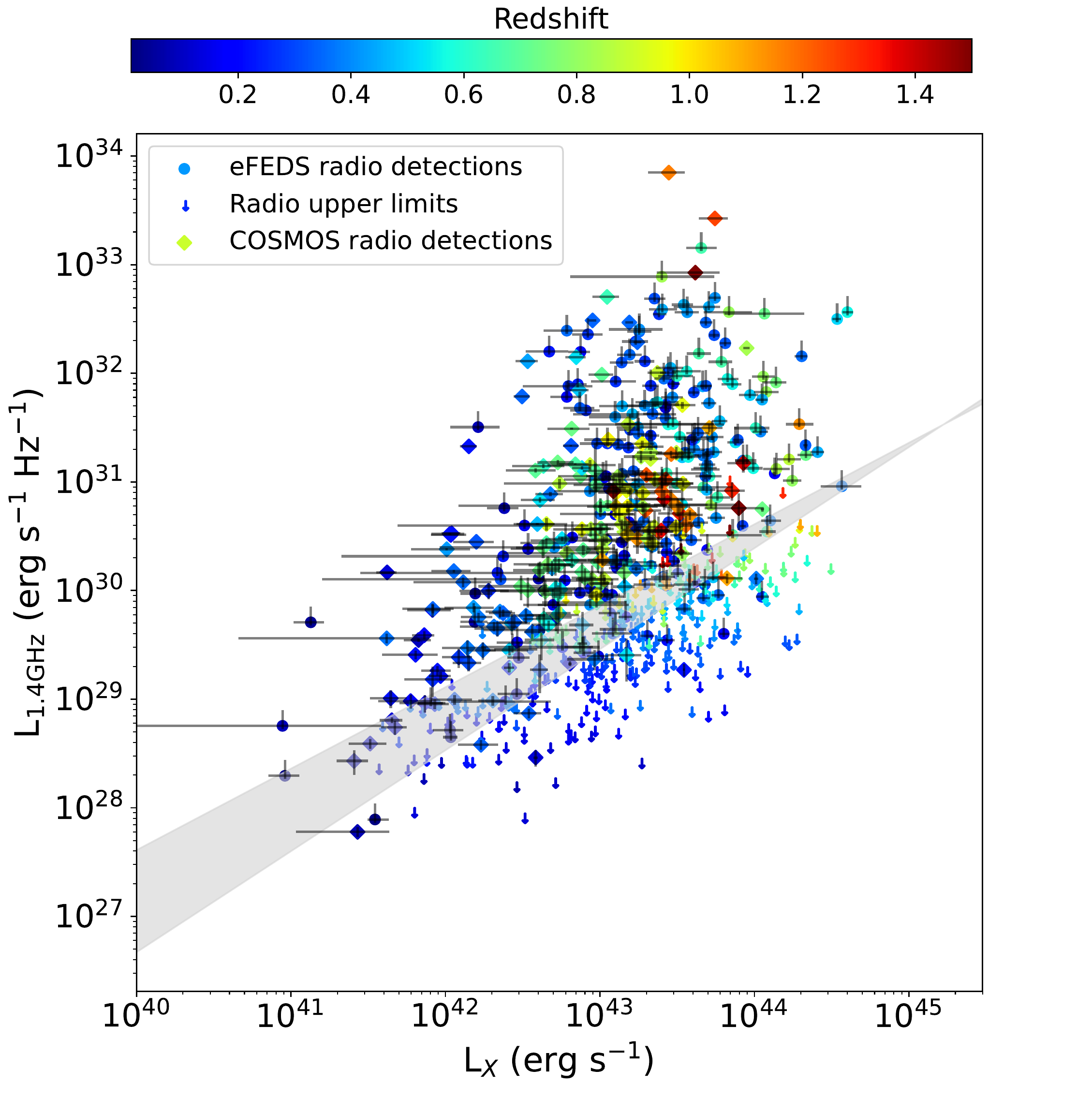}
	\caption{1.4 GHz power of radio galaxies vs. X-ray luminosity of the host cluster for both eFEDS and P20 samples. The colours correspond to the redshift. eFEDS data is represented by circles, while diamonds are COSMOS systems. Downward pointing arrows denote radio upper limits. Bars represent errors in both axes. Errors on $y$-axis are dominated by the assumed uncertainty on the spectral index. The best-fit relation is shown in grey: log$L_R = (0.84 \pm 0.09) \, \log L_X - (6.46 \pm 4.07)$.}
	\label{fig:corrall}
\end{figure}

We compare our 144 MHz sample in eFEDS with a subsample of 137 systems among the 247 COSMOS galaxy groups studied at 1.4 GHz in P20. A further cross-match of our sample with all-sky surveys at this frequency (e.g. NVSS, \citealt{Condon_1998}) is not trivial, due to significant differences in surface brightness sensitivity and resolution. For this reason, the 144 MHz luminosities were converted to luminosities at a frequency of 1.4 GHz assuming $\alpha = 0.8 \pm 0.2$.  The assumed uncertainty on the spectral index dominates on the previous 144 MHz flux error. Combining the two catalogs, we get a total of 364 galaxy clusters and groups that allow us to assess the radio/X-ray correlation using a larger sample. The corresponding $\log L_R$ - $\log L_X$ plot is shown in Fig.~\ref{fig:corrall}.

The distributions of COSMOS and eFEDS clusters and groups are in good agreement. This is confirmed by the two-dimensional Kolmogorov-Smirnov test that, under the null-hypothesis that the two samples are drawn from the same parent distribution, gives $p=0.41$.
This implies that our assumption of a uniform spectral index of $\alpha = 0.8$ for every radio galaxy is valid, although it introduces more scatter in the correlation. Still, a clear trend for more massive groups and clusters hosting more powerful radio sources is seen. This is also supported by the Kendall's $\tau$ test, that for eFEDS$+$COSMOS gives $p < 0.0001$ ($\tau = 0.1331$, $\sigma = 0.0138$). The best-fit relation was estimated exploiting the parametric EM algorithm coded in the AStronomical SURVival statistics package (ASURV, \citealt{Feigelson_2014}), that takes into account different contributions by detections and upper limits. We find log$L_R = (0.84 \pm 0.09) \, \log L_X - (6.46 \pm 4.07)$. This estimate is marginally consistent with the best-fit relations of P20 ($\log L_R = (1.07 \pm 0.12) \times \log L_X - (15.90 \pm 5.13$) and \citet{Pasini_2021b} ($\log L_R = (0.94 \pm 0.43) \times \log L_X - (9.53 \pm 18.19$)), obtained through the same method and applying Bayesian inference, respectively. 

The correlation may imply a link between radiative cooling from the ICM, and the more variable and episodic activity of the AGN. Since the X-ray luminosity is predominantly driven by the cluster or group mass, such a correlation may be produced by massive clusters hosting more massive BCGs, and in turn more massive BHs. In relaxed clusters, the cooling of the ICM is able to efficiently feed the central AGN, leading to higher radio powers \citep{Soker-Pizzolato_2005, Gaspari_2011a}. This is reflected in the well-studied link between the cavity power of systems hosting X-ray bubbles and the luminosity of the cluster cooling region \citep[e.g.,][]{Birzan_2004, Rafferty_2006, Birzan_2017}. \citet{Sun_2009b} also argued that small coronae of X-ray emitting gas in BCGs are able to trigger strong radio outbursts long before cool cores are formed in the host cluster, leading to heating in their surroundings and even preventing their formation, especially in low mass systems. The correlation presented here shows a large scatter, especially at high luminosities. This could be caused, e.g.  by differences in the dynamical states, which we will explore in the next section.


\subsection{Kinetic luminosity and AGN feedback}
\label{sec:kin}

\begin{figure}
	\centering
	\includegraphics[height=27em, width=26em]{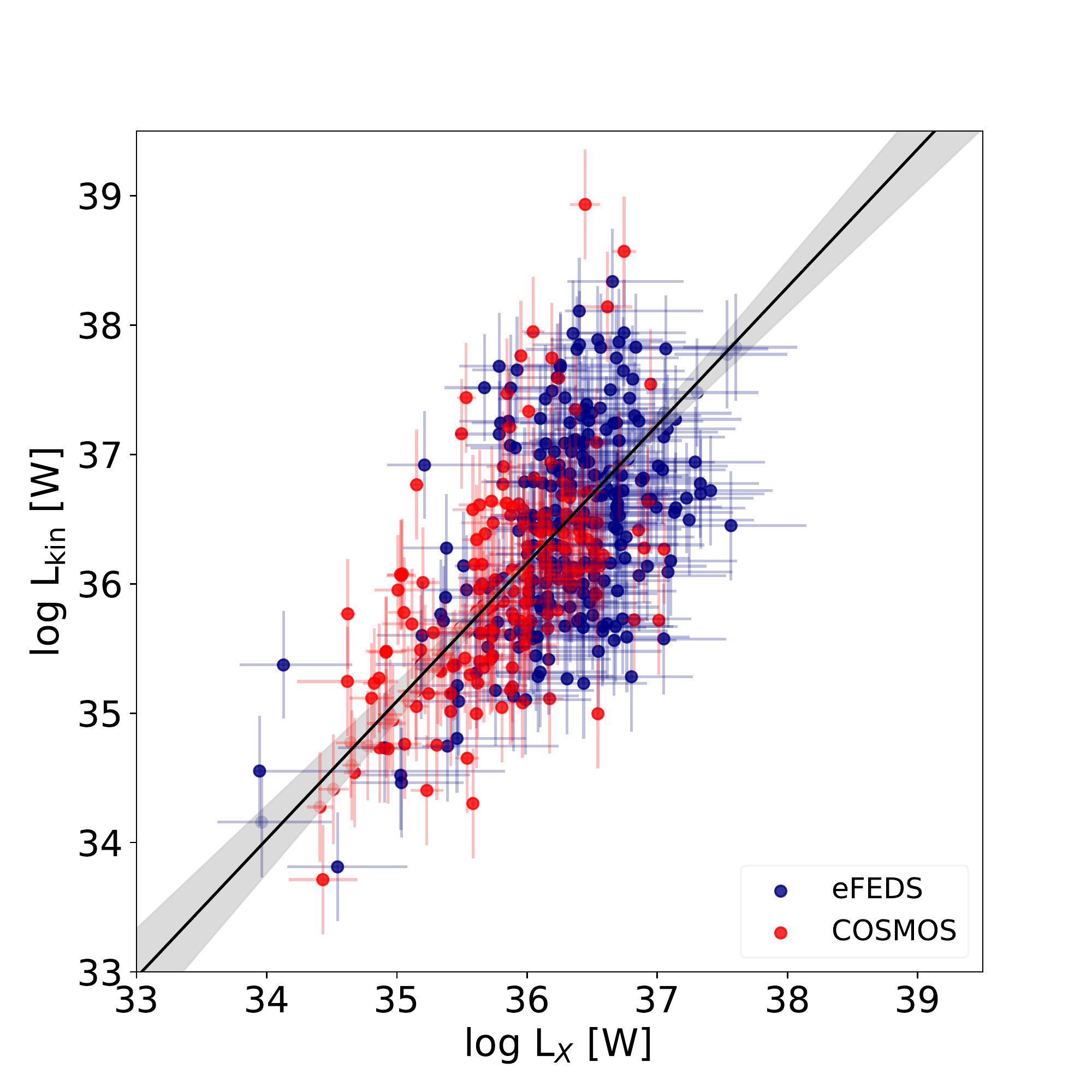}
	\caption{Kinetic luminosity of BCG radio galaxies estimated at 1.4 GHz vs. X-ray luminosity of the host cluster/group for the eFEDS (blue) and P20 (red) samples. The black line represents the best-fit estimated from Bayesian inference: log$L_{\rm kin} = (1.07 \pm 0.11) \ {\rm log} L_X - (2.19 \pm 4.05)$. The grey area indicates 1$\sigma$ errors.}
	\label{fig:kinlum}
\end{figure}

\begin{figure*}
	\centering
	\includegraphics[height=27em, width=26em]{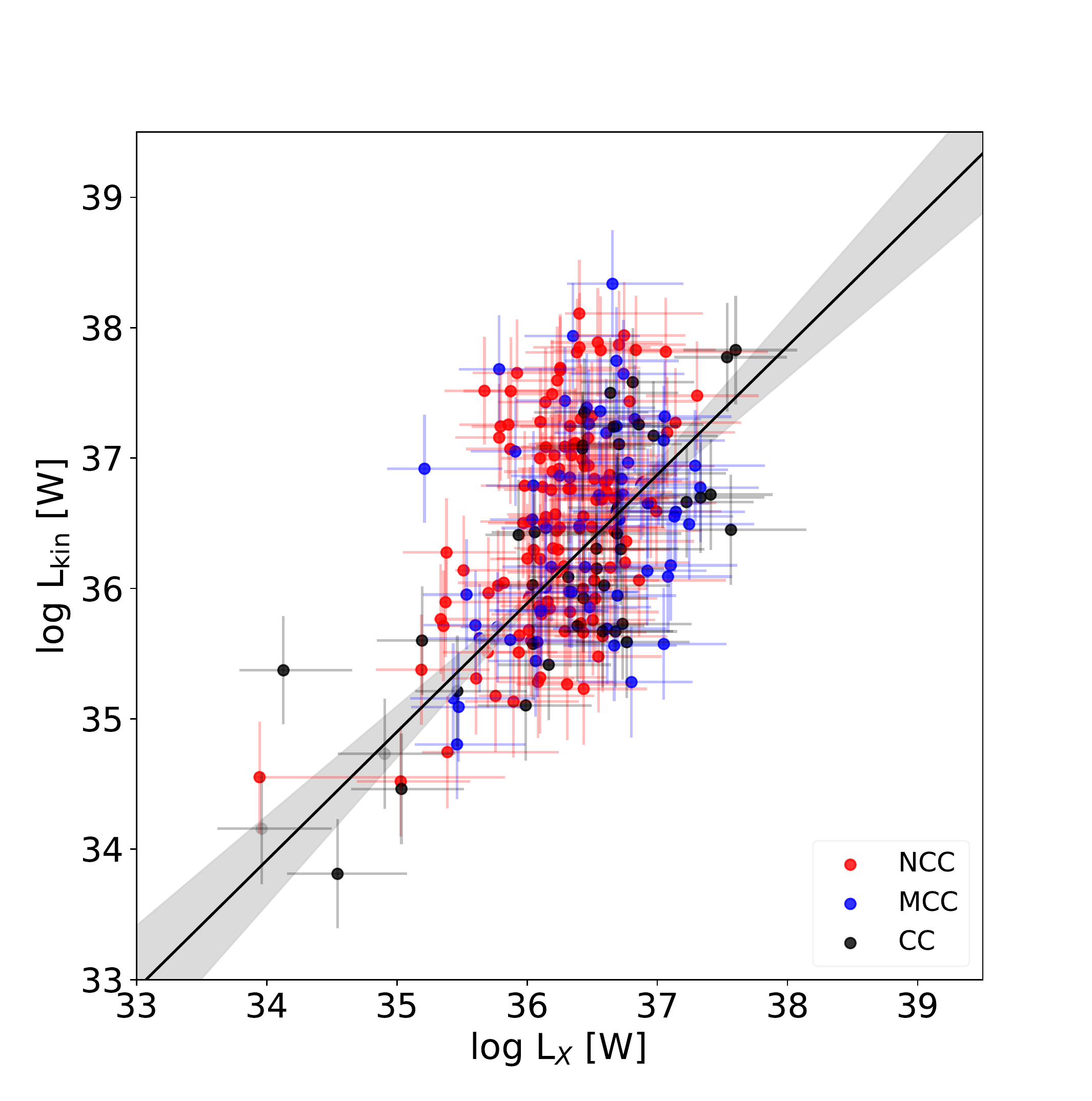}
	\includegraphics[height=27em, width=26em]{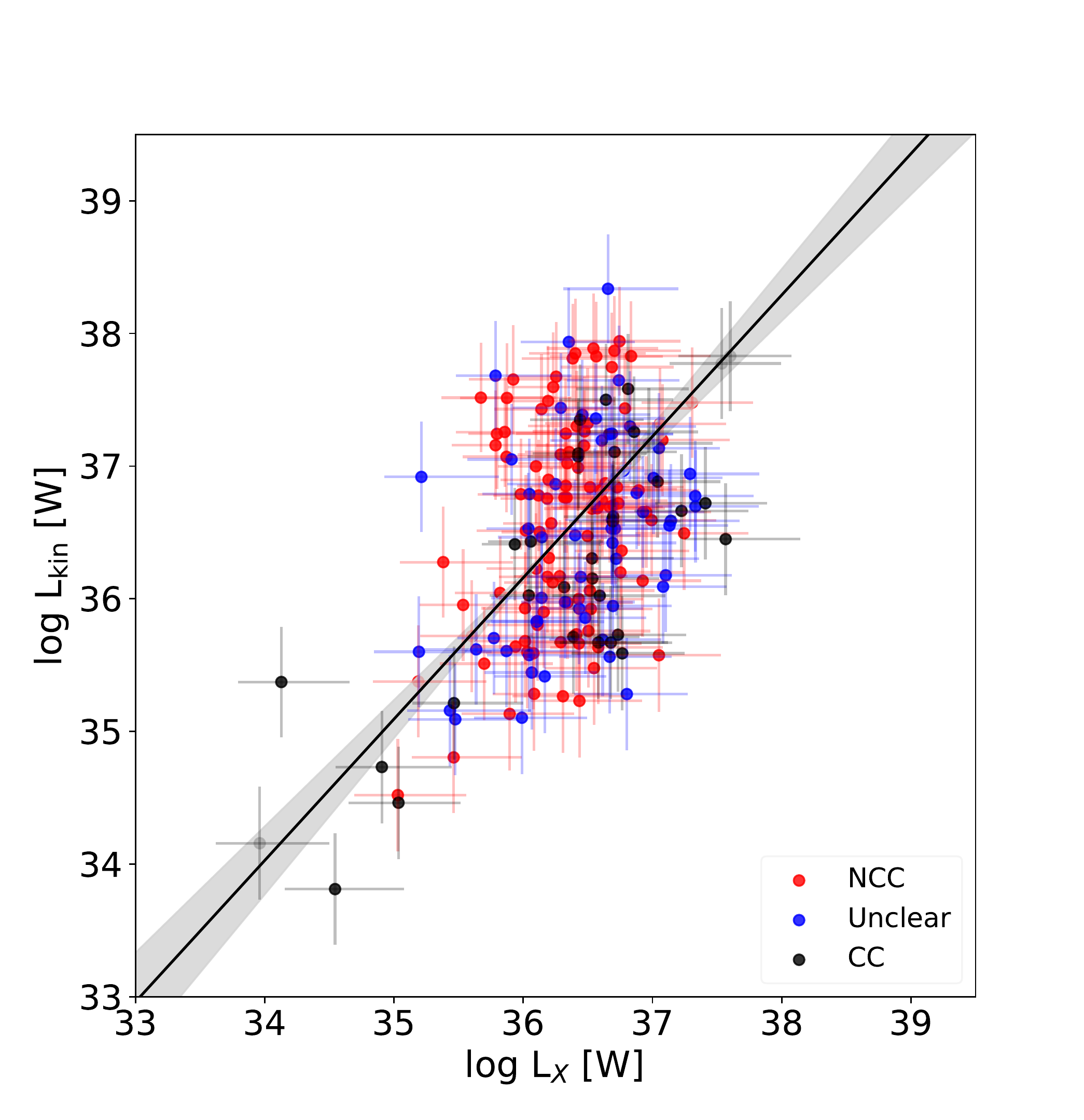}
	\caption{\textit{Left}: Kinetic luminosity of BCG radio galaxies estimated at 1.4 GHz vs. X-ray luminosity of the host cluster/group for the eFEDS sample. Data are classified into non-cool cores (red), moderate cool cores (blue) and cool cores (black) based on the concentration parameter. \textit{Right}: Same of left Panel, with data classified based on the $R_{\rm score}$ (see discussion).}
	\label{fig:classification}
\end{figure*}

The radio luminosity is a measure of the instantaneous radiative loss rate of the radio lobes, and as such is only indirectly related to the energy produced by the AGN through accretion onto the SMBH. For an active source, only a small fraction of the total power supplied to the lobes has been radiated away at any given time, while a much larger fraction is stored in the lobes and a similar amount has been dissipated into the surrounding ICM during the expansion of the jets through the ICM \citep{Willott_1999, Smolcic_2017}. The latter, which we will refer to as kinetic luminosity, is directly linked to the heating of the ICM and contributes to quench the radiative losses of the hot plasma (see Sec.~\ref{sec:intro} for references). 

The relation between the kinetic and radio luminosity has been the subject of ample work \citep[e.g.,][]{Willott_1999, Birzan_2004, Birzan_2008, Cavagnolo_2010, O'Sullivan_2011a, Smolcic_2017}. As thoroughly discussed in \citet{Hardcastle_2019}, there are currently two methods to infer the kinetic luminosity. The first one relies on the identification of X-ray cavities and is affected by assumptions on the cavity age and biased towards small sources in cluster rich environments \citep{Birzan_2012}. The second method relies on a conversion based on a theoretical model and, as such, can lead to unrealistic results if the contribution of source age, environment and redshift to the radio luminosity are not taken into account properly. 
We refer to \citet{Hardcastle_2019} and Appendix A of \citet{Smolcic_2017} for a detailed discussion of this scaling relation. Here, we assume the relation adopted by \citet{Willott_1999} for converting to the 1.4 GHz rest-frame luminosity \citep{Heckman_2014}:

\begin{equation}
    {\rm log} L_{\rm kin, 1.4GHz} = 0.86 \, {\rm log} L_{\rm 1.4GHz} + 14.08 + 1.5\, {\rm log} f_W ,
\end{equation}
where $L_{\rm kin, 1.4GHz}$ is the kinetic luminosity, $L_{\rm 1.4GHz}$ is the luminosity as measured at 1.4 GHz, while $f_W$ is an uncertainty parameter that we assume $f_W = 15$, as estimated by X-ray observations of ICM bubbles in galaxy clusters \citep[e.g.,][]{Merloni_2007, Birzan_2008}. We determine the kinetic luminosity for the radio galaxies of the eFEDS and the P20 sample, and we compare it to the X-ray luminosity within 500 kpc of the host cluster. The result is shown in Fig.~\ref{fig:kinlum}.

In order to infer the relationship between the X-ray and the kinetic luminosity, we applied Bayesian inference on the two samples using the {\ttfamily linmix\footnote{\url{https://github.com/jmeyers314/linmix}}} package \citep{Kelly_2007}. With this tool, we performed a linear fit in the log-log scale in the form:

\begin{equation}
    Y = \alpha + \beta X + \epsilon,
\end{equation}
with $\alpha$ and $\beta$ representing the intercept and the slope, respectively, while $\epsilon$ is the intrinsic scatter of the relation. We find $\alpha =  -2.19 \pm 4.05$, $\beta =  1.07 \pm 0.11$ and $\epsilon = 0.25 \pm 0.05$. We notice that the conversion from radio to kinetic luminosity, which also depends on external factors --- such as the morphology and age of the radio source, the extrapolation of 1.4 GHz fluxes, or the surrounding environment --- and relies on theoretical models, may have introduced artificial scatter into the correlation.

Nevertheless, the plot suggests that in most clusters and groups the heating from the central AGN efficiently counterbalances the ICM radiative losses, as already found in a large number of publications (see references above and \citealt{McNamara-Nulsen_2007, McNamara-Nulsen_2012} for reviews). However, most of these papers take into account the luminosity from within the cooling region, which  is usually defined as the cluster region within which the cooling time of the ICM is shorter than 7.7 Gyr. These usually range between $\sim$50 and $\sim$150 kpc \citep{Birzan_2017}, and their extent can only be estimated through deprojected analysis of the thermodynamical profiles (i.e. temperature, density, cooling time) derived from X-ray observations. The detection of cavities as a tell-tale for AGN heating \citep{McNamara_2000, Birzan_2004} usually requires deep, high-resolution X-ray observations as well. 

The kinetic luminosity--X-ray luminosity relation, estimated through survey data --- albeit with the unprecedented sensitivity of eROSITA --- is able to provide a first insight into the processes of AGN feedback of a large number of clusters and groups. Kinetic and X-ray luminosity act as proxy for mechanical feedback and cooling luminosity, respectively, which together constitute the `parent' correlation usually found in cool core clusters. Nevertheless, here the analysis is performed on all our objects, with no distinction between cool cores and merging clusters. \citet{Main_2017} found that, in their sample of clusters, such a correlation only holds for cool cores. Their classification is based on the central cooling time, determined through \textit{Chandra} observations at 0.004$R_{500}$ by \citet{Hudson_2010}. The eROSITA observations do not yield cooling times at such small cluster-centric radii and we are not able to reproduce the same classification for our objects. Instead we quantify the dynamical state of clusters via the concentration parameter as defined in \citet{Lovisari_2017} and estimated for eFEDS clusters in G21 as:

\begin{equation}
\quad \quad  c_{\rm SB} = \frac{S_B(< 0.1 R_{500})}{S_B(< R_{500})} ,
\end{equation}
where $S_B$ is the surface brightness estimated inside $0.1R_{500}$ for the numerator, and inside $R_{500}$ for the denominator. This parameter is an indicator of the presence of a centrally peaked X-ray surface brightness profile, which correlates with the dynamical state of the cluster. \citet{Lovisari_2017} discusses the use of different thresholds to classify clusters into cool cores and disturbed systems, showing how completeness (i.e. being able to pick all clusters belonging to a given class) and purity (i.e. being able to securely assign clusters to a given class) change depending on the chosen threshold. Here, following the work cited above, we choose to define as non-cool cores (NCC) clusters with $c_{\rm SB} < 0.15$, while cool cores (CC) have $c_{\rm SB} > 0.27$. This classification allows for 100\% purity for both subsamples, albeit completeness goes down to $\sim$53\% for CC and $\sim$75\% for NCC (see \citealt{Lovisari_2017} for more details). Clusters with $0.15 < c_{\rm SB} < 0.27$ cannot be securely categorised, and will be arbitrarily referred to as moderate cool cores (MCC). In the left panel of Fig.~\ref{fig:classification} we show the $L_{\rm kin}$--$L_X$ plot for the eFEDS cluster sample, in which clusters were classified via the concentration parameter. We find that $\sim$53\% of clusters are NCC, $\sim$28\% are MCC and $\sim$19\% are CC. We see no obvious difference in the distribution between the three subsamples. Therefore, the dynamical state of the cluster does not seem to have a large effect on the scatter in the X-ray/radio relation.

As discussed in more details in G21, the concentration acts as an indicator of the presence of a cool core. However, while a relaxed cluster will generally present a cool core, a cool core is not always an indication of relaxation: a merger in its initial stage affects predominantly the cluster outskirts and does not disrupt the cool core  (see e.g. theoretical work by \citealt{Rasia_2015} and \citealt{Biffi_2016}). Therefore, classifying the dynamical state of clusters based on concentration alone is useful to distinguish disturbed objects with low concentration, but does not provide a clear identification of relaxed clusters (see Fig. 9 of G21 and related discussion). For this reason, we perform an alternative classification based on a new morphological parameter first introduced in the same paper, the so-called Relaxation score ($R_{\rm score}$). Since a complete, physical definition of this parameter requires detailed discussions about a number of parameters (see below), we remind to G21 for more insights, and here we just provide a brief description. The $R_{\rm score}$ combines a number of morphological parameters usually determined for galaxy clusters, such as concentration, central density, ellipticity (ratio between minor and major axes of the cluster), cuspiness (slope of the density profile at a given radius) and others. The resulting $R_{\rm score}$ provides a more clear indication, with respect to concentration alone, of the dynamical state of a cluster. In particular, the $R_{\rm score}$ should be higher for relaxed objects, that show large concentration, central density, ellipticity and cuspiness. On the other hand, the same parameter should decrease in disturbed clusters.

Following the discussion in G21, we define as relaxed objects with $R_{\rm score} >$ 0.0137. The results of this alternative classification are shown in the right panel of Fig. \ref{fig:classification}. We only plot clusters for which a proper estimate of the $R_{\rm score}$ was feasible in G21. Objects classified as relaxed through the $R_{\rm score}$ and as CC through the concentration are generally referred to as CC, those with low $R_{\rm score}$ and concentration are NCC, while clusters with high concentration (same threshold used for left panel) but $R_{\rm score} <$ 0.0137 are labeled as unclear. We remind to G21 for discussions and comparisons between different classifications, while in this work we focus on the correlation.

Even when introducing a more accurate parameter such as the $R_{\rm score}$, there is still not a clear distinction in the distribution between cool cores and merging objects, as instead found e.g. in \citet{Main_2017}.
Furthermore, it is not clear how such a relation could be present in disturbed systems. In these objects, the cooling of the ICM is slow and BCGs often hard to identify. Morphological parameters have been widely used to determine the dynamical state of clusters, but a more secure classification based on central cooling time may be more useful to understand in which clusters a connection of AGN and cooling ICM can ensue. A possibility is that the link between AGN and their environment could be produced, even in disturbed objects, by rapidly-cooling coronae permeating the host galaxy \citep{Sun_2007, Gastaldello_2008, Sun_2009b}. This idea has been already suggested for NCC hosting radio AGN, such as A2028 \citep{Gastaldello_2010}. It is also plausible that small, low-entropy regions of the cluster core such as cool core remnants \citep{Rossetti_2010} could affect the AGN, leading to the observed relation. Another possibility is that NCC do not belong, in fact, to the correlation. To test this, we checked the scatter of the correlation after applying Bayesian inference only on CC. If NCC are not part of the correlation, the scatter of the data should decrease when fitting only CC. We find $\epsilon = 0.17 \pm 0.10$, consistent within errors with the previous estimate. Nevertheless, the uncertainty increases because of the relatively small number of CC, and further analyses exploiting larger samples are needed to investigate this further.


\section{Conclusions}
\label{sec:conclusions}

We make use of eROSITA (X-ray) and LOFAR (radio) observations of the eFEDS field in order to investigate radio galaxies hosted in BCGs. Our results can be summarised as follows:

\begin{itemize}
    \item Our sample yields 227 detections and 248 upper limits in the redshift range $0.01 < z < 1.3$ and luminosity range $10^{22}$ -- $10^{27}$ W Hz$^{-1}$ at 144 MHz. The remaining 67 clusters were excluded from the analysis to avoid contamination by missclassified AGN (see Sec. \ref{sec:construction}). The radio detection rate is $\sim$48\%, which is lower than in other samples of well-studied groups and clusters. 

    \item BCGs hosting radio-loud AGN mostly ($\sim$84\%) lie within 50 kpc from the cluster centre. BCGs that are more offset tend to have lower levels of radio emission, or lie below our detection threshold.
    
    \item As already argued in previous works, larger radio galaxies are usually more powerful. However, we note that a relevant selection effect is present in our sample, since we lack large, low-power radio sources because of surface brightness limitations. We see no correlation of the cluster's central ($R = 0.02R_{500}$) density with the LLS, suggesting that the luminosity is a better predictor for the size of the radio galaxy.
    
    \item We studied the relation between the 144 MHz radio galaxy power and the host cluster X-ray luminosity measured within 500 kpc from the cluster centre, finding a positive correlation. Because of the large number of upper limits, we relied on statistical tests, such as the partial correlation Kendall's $\tau$ test and the scrambling test, to show that the correlation is not produced by selection effects in the radio band.
    
    \item Converting the 144 MHz power of radio galaxies to 1.4 GHz, we compared our results with the correlation between the X-ray luminosity and the 1.4 GHz power of a COSMOS galaxy groups sample first investigated by \citet{Pasini_2020}. We found that the two samples are in good agreement based on a Kolmogorov-Smirnov test that, under the null-hypothesis that the samples are drawn from the same parent distribution, gives $p=0.41$. We estimated a best-fit relation log$L_R = (0.84 \pm 0.09) \, \log L_X - (6.46 \pm 4.07)$.
    
    \item We converted the radio powers of radio galaxies to kinetic luminosities, making use of widely used scaling relations. Comparing the kinetic luminosity to the X-ray luminosity within 500~kpc from the cluster centre, we found that in most objects the ICM's radiative losses are efficiently counterbalanced by heating supplied from the central AGN. We derived the best-fit relation applying Bayesian inference, obtaining ${\rm log} L_{\rm kin} = (-2.19 \pm 4.05) + (1.07 \pm 0.11)\, {\rm log} L_X + (0.25 \pm 0.05)$.
    
    \item We classified eFEDS clusters into disturbed and relaxed objects based on two different parameters, concentration and Relaxation score (see Sec.~\ref{sec:kin} for a definition). We could see no significant differences in the $L_{\rm kin}-L_X$ relation between the subsamples.
    
\end{itemize}

Future prescriptions of radio-mode AGN feedback in simulations need to be able to recover the properties described in this paper. In addition to massive halo gas fractions, entropy slopes, and galaxy properties, they need to recover radio luminosities as a function of the host cluster properties. With the new all-sky X-ray surveys, a correlation between the cluster X-ray luminosity and the BCG radio power can be used to probe AGN feedback across a larger range of host masses and to control for the effect of other observables. Particularly, the synergy between eRASS (Bulbul et al. in prep.) and the LOFAR Two-Metre Sky Survey (LoTSS, \citealt{Shimwell_2017}), as well as the forthcoming LOFAR LBA Sky Survey (LoLSS, \citealt{deGasperin_2021}), will provide samples of thousands of clusters and groups for which the interplay between the AGN and the ICM can be investigated.

\section*{Acknowledgements}

TP thanks Philip Best for useful comments.
TP is supported by the BMBF Verbundforschung under grant number 50OR1906.
MB acknowledges support from the Deutsche Forschungsgemeinschaft under Germany's Excellence Strategy - EXC 2121 "Quantum Universe" - 390833306. 
DNH acknowledges support from the ERC through the grant ERC-Stg DRANOEL n. 714245. 
AB acknowledges support from the VIDI research programme with project number 639.042.729, which is financed by the Netherlands Organisation for Scientific Research (NWO). FG acknowledges support from INAF mainstream project ‘Galaxy Clusters Science with LOFAR’ 1.05.01.86.05.
RJvW acknowledges support from the ERC Starting Grant ClusterWeb 804208.
WLW  acknowledges support from the CAS-NWO programme for radio astronomy with project number 629.001.024, which is financed by the Netherlands Organisation for Scientific Research (NWO).
This work is based on data from eROSITA, the soft X-rays instrument aboard SRG, a joint Russian-German science mission supported by the Russian Space Agency (Roskosmos), in the interests of the Russian Academy of Sciences represented by its Space Research Institute (IKI), and the Deutsches Zentrum für Luft- und Raumfahrt (DLR). The SRG spacecraft was built by Lavochkin Association (NPOL) and its subcontractors, and is operated by NPOL with support from the Max Planck Institute for Extraterrestrial Physics (MPE).
The development and construction of the eROSITA X-ray instrument was led by MPE, with contributions from the Dr. Karl Remeis Observatory Bamberg \& ECAP (FAU Erlangen-Nuernberg), the University of Hamburg Observatory, the Leibniz Institute for Astrophysics Potsdam (AIP), and the Institute for Astronomy and Astrophysics of the University of Tübingen, with the support of DLR and the Max Planck Society. The Argelander Institute for Astronomy of the University of Bonn and the Ludwig Maximilians Universität Munich also participated in the science preparation for eROSITA.
The eROSITA data shown here were processed using the eSASS/NRTA software system developed by the German eROSITA consortium.
LOFAR data products were provided by the LOFAR Surveys Key Science project
(LSKSP; https://lofar-surveys.org/) and were derived from observations with the International LOFAR Telescope (ILT). LOFAR (van Haarlem et al. 2013) is the Low Frequency Array designed and constructed by ASTRON. It has observing, data processing, and data storage facilities in several countries, that are owned by various parties (each with their own funding sources), and that are collectively operated by the ILT foundation under a joint scientific policy. The efforts of the LSKSP have benefited from funding from the European Research Council, NOVA, NWO, CNRS-INSU, the SURF Co-operative, the UK Science and Technology Funding Council and the J\"ulich supercomputing centre.

\bibliographystyle{aa.bst}
\bibliography{bibliography, erosita, radio}

\begin{appendix}
\section{Examples of interesting systems}

\begin{figure*}
	\centering
	\includegraphics[scale=0.65]{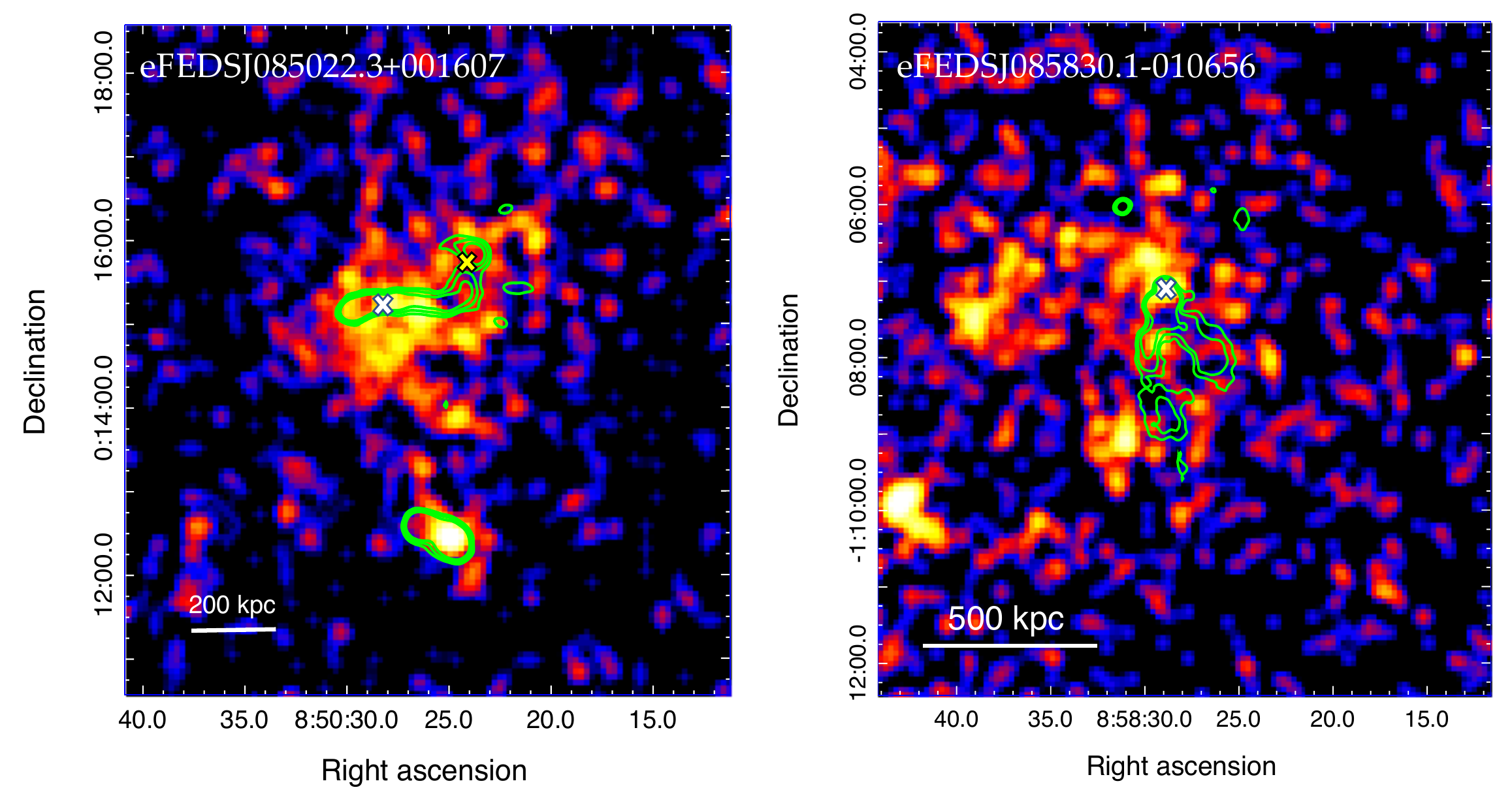}
	\caption{eROSITA 0.2-2.3 keV images of eFEDSJ085022.3+001607 (left panel) and eFEDSJ085830.1-010656 (right panel), smoothed with a 3$\sigma$ gaussian filter. LOFAR 144 MHz contours at 3,6,12,24 $\cdot \ rms$ (local) are in green. The white cross represents the cluster X-ray peak, while the yellow cross is the BCG position. For eFEDSJ085830.1-010656, the BCG is coincident with the X-ray peak.}
	\label{fig:images1}
\end{figure*}

\begin{figure*}
	\centering
	\includegraphics[scale=0.65]{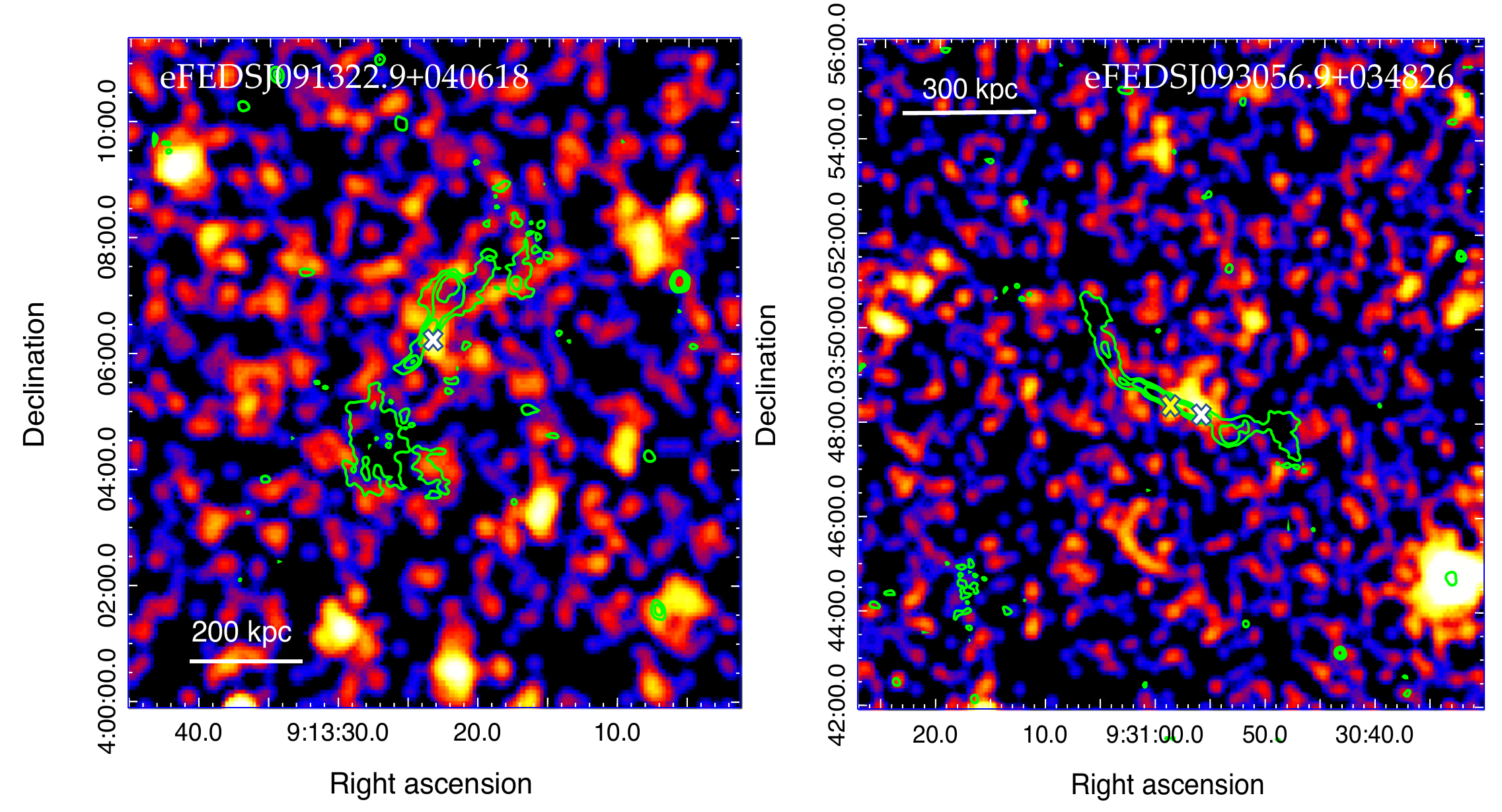}
	\caption{eROSITA 0.2-2.3 keV images of eFEDSJ091322.9+040618 (left panel) and eFEDSJ093056.f9+034826 (right panel), smoothed with a 3$\sigma$ gaussian filter. LOFAR 144 MHz contours at 3,6,12,24 $\cdot \ rms$ (local) are in green. The white cross represents the cluster X-ray peak. For eFEDSJ091322.9+040618, the BCG is coincident with the X-ray peak.}
	\label{fig:images2}
\end{figure*}

\begin{table*}
	\centering
	\caption{X-ray observables and BCG radio power for 4 relevant eFEDS clusters}
	\begin{tabular}{c c c c c c}
		\hline\hline
		Name & $z$ & $kT^a$ [keV] & $L_{\rm bol}^a$ [10$^{43}$ erg s$^{-1}$] & $c_{\rm SB}$ & $L_R^b$ [10$^{24}$ W Hz$^{-1}$] \\
		\hline
		eFEDSJ085022.3+001607 & 0.196 & $3.1 \pm_{0.7}^{1.1}$ & $2.7 \pm_{0.9}^{1.3}$ & 0.02 & $4.1 \pm 0.2$\\
		eFEDSJ085830.1-010656 & 0.224 & $2.1 \pm_{0.8}^{1.7}$ & $2.1 \pm_{0.4}^{0.5}$ & 0.13 & $25.0 \pm 1.0$\\
		eFEDSJ091322.9+040618 & 0.088 & $0.45 \pm_{0.17}^{0.29}$ & $4.1 \pm_{0.9}^{1.2}$ & 0.04 & $0.74 \pm 0.05$\\
		eFEDSJ093056.9+034826 & 0.09 & $0.61 \pm_{0.27}^{0.75}$ & $2.7 \pm_{0.9}^{1.2}$ & 0.21 & $0.77 \pm 0.02$\\
		\hline\hline
	\end{tabular} \\
	Notes: $^a$: estimated within 500 kpc. $^b$: 144 MHz luminosity of the BCG.
	\label{tab:4sys}
\end{table*}

The high flux sensitivity and spatial coverage of eROSITA and LOFAR at their respective frequencies allows for interesting comparisons. In the past, the combination of X-ray and radio observations of galaxy clusters and of their BCGs have led to a significant improvement in the understanding of the thermal and non-thermal processes in these environments (e.g. \citealt{Gitti_2010, Kolokythas_2018, Botteon_2020}; see Sec.~\ref{sec:intro} for more references and reviews).

We used the eROSITA and LOFAR observations to look for systems showing interesting morphologies and signs of possible interaction between the ICM and the central AGN. In this section, we present four among the most interesting examples of such clusters. We focus on AGN emission only, while diffuse emission more directly associated with the ICM and clusters dynamical state will be presented in a forthcoming paper (Hoang et al. in prep.). Table \ref{tab:4sys} summarises the main properties of these systems.

\subsubsection{eFEDSJ085022.3+001607}

eFEDSJ085022.3+001607 (left panel of Fig.~\ref{fig:images1}) is located at a redshift of $z=0.196$ (spectroscopic). The strongly elliptical and irregular morphology of the X-ray emission and low concentration ($c_{\rm SB}=0.02$) suggest that this cluster is disturbed. The BCG hosts an elongated, head-tail shaped radio galaxy (major axis $\sim$ 500 kpc), with a 144 MHz luminosity of $L_R = (4.1 \pm 0.2) \times 10^{24}$ W Hz$^{-1}$. The AGN lies at $\sim$ 150 kpc from the X-ray peak. Surface brightness discontinuities that coincide with the lobes of the radio galaxy are detected in the X-ray image. However, the relatively low resolution does not reveal any ICM cavities, which however have never been detected around head-tails. The shape of the non-thermal emission follows that of the hot plasma, with the jet extending towards the East through the X-ray ripple. Meanwhile, the expansion in the opposite direction appears frustrated.

\subsubsection{eFEDSJ085830.1-010656}

The irregular morphology and low concentration ($c_{\rm SB}=0.13$) of eFEDSJ085830.1-010656 (right panel of Fig.~\ref{fig:images1})  leads us to classify it as a non cool core. The BCG hosts a wide angle tail radio galaxy with two tails departing in the S and SW directions for $\sim$ 250 kpc each. The tails are expanding into a lower-density region within the group. Deeper X-ray observations are needed to study the ICM emission of this group due to its low surface brightness and relatively high redshift.

\subsubsection{eFEDSJ091322.9+040618}

eFEDSJ091322.9+040618 (left panel of Fig.~\ref{fig:images2}) is a low-redshift ($z=0.088$, spectroscopic) galaxy group classified as a disturbed cluster due to its irregular shape and low concentration ($c_{\rm SB} =0.04$). The radio galaxy extends for more than 200 kpc along the NW-SE axis. The lobes are expanding into the SE and NW directions following the hot plasma. Diffuse emission with unclear origin is detected in the SE direction, correspondingly to a low surface brightness region, extending for $\sim$ 150 kpc.  

\subsubsection{eFEDSJ093056.9+034826}

eFEDSJ093056.9+034826 (right panel of Fig.~\ref{fig:images2}) is a galaxy group located at $z=0.09$ (photometric). The elliptical shape and relatively high concentration ($c_{\rm SB}=0.21$) classify it as a moderate cool core. The BCG hosts a double-lobe elongated radio galaxy with a major axis of $\sim$ 600 kpc and $L_R = (7.7 \pm 0.2) \times 10^{23}$ W Hz$^{-1}$. The long lobes ($\sim 300$ kpc) of the central radio galaxy are extending far beyond the X-ray bright core of the group. The low X-ray flux of this group makes it difficult to identify depressions in the surface brightness.

\end{appendix}

\label{lastpage}
\end{document}